\documentstyle[12pt,aaspp4]{article}

\lefthead{Reid}
\righthead{Globular cluster distances}
\begin {document}
\title{ Younger and brighter - New distances to globular clusters based on
Hipparcos parallax measurements of local subdwarfs}
%\footnote
%{Based in data from the ESA Hipparcos astrometry satellite}

\author {I. Neill Reid}
%\footnote
%{ Visiting Research Associate, Observatories of the Carnegie Institution of
%Washington}
\affil {Palomar Observatory, 105-24, California Institute of Technology, Pasadena, CA 91125,
e-mail:  inr@astro.caltech.edu}
\begin {abstract}

We have used new parallax measurements, obtained by the Hipparcos satellite, of
fifteen nearby, metal-poor stars to re-define the subdwarf main-sequence.
All of these stars have parallaxes determined to an accuracy of at least 12 \%.
Comparing these measurements against previous ground-based data for nine stars
reveals a systematic offset of 5 \%, in the sense that the Hipparcos
parallaxes are smaller (i.e. the inferred distances are larger).
The availability of the Hipparcos observations expands the local subdwarf sample to the extent
that we can separate the stars by abundance into intermediate ([Fe/H] $\sim$ -1.4) and
extreme ([Fe/H] $\sim -2$) subsets. Main-sequence fitting techniques are then used
to match stars of the appropriate abundance range to the colour-magnitude
diagrams of the seven globular clusters M5, NGC 6752, M13, M15, M92, M30 and M68. 
We derive respective distance moduli of 14.45, 13.17, 14.48, 15.38, 14.93, 14.95 and
15.29 magnitudes, with formal uncertainties of $\pm 0.1$ magnitude.
The metal-poor systems M68, M15 and M30 have moderate foreground reddening, and
varying E$_{B-V}$ by $\pm 0.02$ magnitudes can change the derived distances by up
to $\pm 7 \%$. With the exception of NGC 6752, however, our derived distances exceed previous
estimates, particularly in the case of the four [Fe/H]$\sim -2.1$ globulars, where
our distance moduli are $\sim 0.3$ magnitudes higher than the current standard values.
We discuss briefly how these findings affect the RR Lyrae distance scale, isochrone-based estimates 
of the age of globular clusters and our picture of the early stages of star formation
in the Galaxy.  We note that our results go
some way towards reconciling the apparent contradiction between the cluster ages
and recent determinations of the Hubble constant.

\end {abstract}


\section {Introduction }

Globular clusters constitute a vital rung on the extragalactic distance ladder.
Not only do they provide a means of calibrating the absolute magnitudes of
secondary distance indicators, such as short-period miras and RR Lyraes, 
but also comparisons between
their colour-magnitude diagrams and the predictions of stellar evolutionary
models can be used both to probe the  early star-formation history of the
Galaxy and to set a lower limit on the age of the universe. Both
issues are controversial at present. In the first case, most isochrone-based
colour-magnitude diagram analyses derive ages of from 14 to 18 Gyrs
for the majority of globular clusters, (Chaboyer, Demarque \& Sarajedini, 1996; 
Sandquist et al, 1996 - hereinafter S96). Bolte \& Hogan (1995 - BH95), in particular,
have argued that matching the best cluster data against theoretical isochrones
leads to age estimates of 15.8$\pm$2.1 Gyrs.
These timescales are in conflict with recent estimates of the expansion age of 
the Universe. The ``mid-term" value of $73 \pm 10$ km s$^{-1}$ Mpc$^{-1}$ derived for
the Hubble constant by the HST key project team (Freedman, Madore \& Kennicutt, 1997)
implies an age of only 9 to 12 Gyrs when interpreted in standard $\Lambda = 0$ 
cosmological frameworks (Freedman, priv. comm.), although the long distance-scale 
value of H$_0 = 50$ to 55 km s$^{-1}$ Mpc$^{-1}$ favoured by Tammann \& Sandage (1996)
is formally consistent with the stellar chronometers.

In the second case, there is a long-standing discrepancy between the distance to the 
Large Magellanic Cloud that one derives using the RR Lyrae absolute magnitude calibration 
and that computed based on the Cepheid scale. The traditional calibration of the 
latter period-luminosity relation, based on Galactic Cepheids, leads to an
LMC distance modulus of 18.57 (Feast, 1997). A small number of the older LMC clusters
with extensive RR Lyrae populations have been monitored photometrically, allowing
the determination of accurate mean magnitudes. In particular, five clusters
with [Fe/H]=-1.8$\pm$0.1 (from Walker, 1992 and Reid \& Freedman, 1994)
give a mean magnitude, corrected for foreground
reddening, of $<V_0> = 18.98 \pm 0.03$ for LMC RR  Lyrae variables. If one
adopts the absolute magnitude calibration given by statistical parallax 
analysis of the halo RR Lyraes in the Solar Neighbourhood, 
$<M_V> = 0.71 \pm 0.12$ mag, with little dependence on abundance (Layden et al, 1996)
then one derives an LMC modulus of only 18.3 magnitudes. 

Baade-Wesselink analyses of 
Galactic field stars, however, suggest that there is a significant correlation between
luminosity and abundance, although there is as yet no clear agreement as to the
exact correlation. Carney et al (1992a) derive a shallow slope of only 0.15 mag. dex$^{-1}$
in M$_V$ and a zeropoint of M$_V$=1.01 at [Fe/H]=0.0; Fernley (1994) finds a slope
of 0.21 and a zeropoint of M$_V$ = 0.97 mag; Cacciari, Clementini \& Fernley (1992)
previously estimated a similar slope of 0.2, but a fainter zeropoint, M$_V$=1.04; and
Sandage (1993) derives a steeper slope, 0.3 mag. dex$^{-1}$, and a brighter
zeropoint, M$_V$=0.94 mag. Finally, Feast (1997) has used inverse regression
to re-analyse the Fernley sample, and drives a yet-steeper slope, 0.37 mag dex$^{-1}$,
and the faintest zeropoint, M$_V$=1.13. Combined with the mean observed magnitude, 
$<V_0> = 18.98$ mag at [Fe/H]=-1.8, these five different calibrations lead to LMC distance
moduli estimates of, respectively, 18.24, 18.39, 18.30, 18.58 and 18.52 magnitudes.
The last two estimates are in acceptable agreement with the traditional Cepheid
values and with the older distance estimate based on SN 1987A (Panagia et al, 1991),
although Gould (1995) has recently advocated (m-M)$_0$=18.37 mag for the latter, matching the
shorter distance scale. \footnote
{Feast \& Catchpole note that Sonnenborn et al. (preprint), argue that a better
estimate is 18.46 $\pm$ 0.10 mag., with possible systematic uncertainties
still to be taken into account.}
However, Feast \& Catchpole (1997) have recently re-calibrated the
zeropoint of the Cepheid period-luminosity relationship using Hipparcos parallax estimates
of Galactic variables, and now propose a true LMC modulus of 18.7
magnitudes, restoring the disagreement between the Cepheid and RR Lyrae-based distance
scales.

As Sandage (1986) has emphasised, 
both cluster age-determinations and the calibration of the absolute magnitude of 
horizontal branch stars rest on the accuracy of the distances to
individual globular clusters. Those distances can be estimated either
empirically, by main-sequence fitting - matching the observed cluster
colour-magnitude diagram against the main-sequence defined by stars
of the appropriate metallicity which have accurate trigonometric parallax
measurements - or in a semi-empirical fashion, matching theoretical isochrones,
calibrated against the local subdwarfs, to
the observed cluster sequences. The latter
technique depends on the availability of accurate bolometric corrections and
temperature-colour calibrations to transform from the (Luminosity, temperature) plane 
to the observed (M$_V$, colour) plane. Generally, most previous studies have
agreed in deriving cluster ages of 14 or more Gyrs from either technique. A notable
exception is the analysis by Mazzitelli, D'Antona and Caloi (1995), based on
updated physical parameters, which lead to more luminous horizontal-branch 
stellar models. Coupled with a new treatment of convection, which 
reduces the turnoff age for a given luminosity,
they derive younger ages (12-14 Gyrs) for the metal-poor clusters.

The primary observational method for deriving cluster distances is main-sequence fitting 
against subdwarfs of known absolute magnitude and colour.
Since globular clusters are significantly more metal-poor than
the Sun, the calibrating parallax stars must be drawn from the Galactic halo
population, which constitutes only $\sim 0.15 \%$ of the local stars. While
there are a few M-type subdwarfs (such as Kapteyn's star) within 10
parsecs of the Sun, the nearest F and G subdwarfs lie at distances of 40-50 parsecs. 
As a result, until recently there were scarcely a dozen stars whose distances
were known to sufficient precision to allow their use as calibrators. Those stars,
moreover, span an abundance range of well over 1 dex, requiring the application
of differential colour corrections (derived from theoretical isochrones) to
adjust each star to an appropriate mono-metallicity sequence 
before matching the colour-magnitude data for an individual cluster.
The required corrections can be substantial:
for example, adjusting the observations of HD 23439A ([Fe/H] = -1.02) to match
the Bergbusch \& Vandenberg (1992 - BV92)
[Fe/H]=-2.0 isochrone demands a correction of $\sim$-0.1 magnitudes to the 
(B-V) colour (cf S96, Table 9).\footnote
{Note that to be physically correct, this procedure should preserve mass - that is,
both the colour and the absolute magnitude of a given subdwarf should be adjusted to
those appropriate to a star of the equivalent mass at the abundance of the cluster.
The M$_V$ corrections are not usually applied in practice.} These corrections, moreover,
are model dependent. D'Antona, Caloi and Mazzitelli (1997 - DCM) derive 
a colour-shift of ${\Delta (B-V)}\over{\Delta [Fe/H]}$ of 0.056 mag, and,
using the same small sample of calibrators as BH95, derive
significantly larger distances to metal-poor clusters like M92. Finally, one
should also bear in mind the fact that cluster abundances are usually measured
using red giant stars, while the field subdwarfs are (obviously) unevolved.

The successful completion of the Hipparcos astrometric satellite project
and the publication of the Hipparcos catalogue (ESA 1997)
has improved markedly the prospects of obtaining accurate distance calibration
to globular clusters. Not only has the
Hipparcos team provided more accurate parallaxes for the ``classical"
subdwarf calibrators, but the sample of metal-poor stars has been increased
to the point at which one can isolate sufficient stars within a limited 
range of abundance to define a main-sequence without recourse to significant
colour corrections. We have obtained Hipparcos data for a sample of proper
motion stars which includes fifteen subdwarfs with abundances in the
range -2.2 $\ge$ [Fe/H] $\ge$ -1.25, and we have combined these observations
with data for three additional stars with accurate ground-based astrometry 
to derive distances to seven globular clusters. In six cases, our
derived distance moduli are higher than currently accepted values, 
particularly for the metal-poor clusters. The following section presents the 
new parallax data; section 3 describes the main-sequence fitting and the
distance derived for each cluster; and section 4 discusses our conclusions.

\section {The subdwarf calibration}

The Hipparcos satellite was launched with the aim of obtaining
accurate positions, parallaxes and proper motions for over 110,000 relatively bright
stars distributed over the whole sky.
Despite the problems introduced by the satellite's failure to leave a highly-elliptical
orbit, the separate consortia completed their data reduction in late 1996, 
achieving the goal of milliarcsecond-precision astrometry. The reduction 
procedures are discussed by Kovalevsky et al (1995), while Perryman et al (1995)
describe the tests carried out to establish the accuracy of the resultant
parallax measurements. Part of the
Hipparcos programme consisted of observations of a substantial fraction of the
Lowell proper motion stars (Giclas, Burnham \& Thomas, 1971) brighter than 
V=12 magnitude, a proposal submitted by the author in 1982. Some 2500 stars
were observed, including five ``classical" subdwarfs (S96, Table 9) 
together with a number of
other well-known subdwarfs, such as HD 19445 and HD 84937. In addition, 
since the inception of the Hipparcos project, 
many of the F, G and early-K stars in the Lowell 
survey have been observed spectroscopically and photometrically by
Carney, Latham and Laird (Carney et al, 1994: CLLA); these observations have 
added several stars to the small sample of known, nearby metal-poor subdwarfs
with accurate photometry and abundance estimates. 

Seven hundred and forty-five of the 1447 stars in the CLLA catalogue 
have Hipparcos astromtery. The complete set of observations offers insight 
into both the detailed structure of the
lower main-sequence and the kinematics of the higher velocity stars in the
disk, and these matters will be discussed in detail elsewhere (Reid, in preparation).
In the present paper we concentrate on the metal-poor stars. One hundred and six of
the CLLA stars observed by Hipparcos have [Fe/H]$<$-1.25, but only thirty-five of these
stars have parallaxes determined to a precision ($\sigma_\pi \over \pi$) of 20 \% or
better. The relevant data for these stars are listed in Table 1, where we have
divided the sample into two groups: fifteen stars with parallaxes measured to a precision of
at least 12 \%, which represent our primary calibrators, and 20 stars with
less precise parallax data. The
photometry and abundance estimates are taken directly from Table 6 of CLLA.
Eleven stars are listed as having significant foreground reddening by CLLA, and
we have corrected the magnitudes and colour appropriately. Six stars are found to
be single-lined spectroscopic binaries by CLLA, and a further 8 stars are
candidate binaries based one either radial velocity data or the Hipparcos astrometry.
These stars are identified in Table 1.
Twenty-six stars from the complete sample are listed in 
the recently-published Yale General Catalogue of Trigonometric
Parallaxes (van Altena, Truen-liang Lee \& Hoffleit, 1995 - hereinafter, YPC),
and Table 1 lists YPC data for three ``classical" subdwarfs which were not 
amongst our Hipparcos observations. However, before comparing these astrometric data 
against previous measurements, it is
important to consider the appropriate method of correcting for the 
bias known to be present in such observations.

\subsection {Systematic corrections to parallax measurements}

Table 1 lists the absolute magnitudes, M$_V$, derived directly from the parallax data. 
These represent the most accurate estimate of the visual luminosity of each star.
However, when combining the individual measurements in any statistical analysis,
one must take into account systematic bias towards overestimating parallax
measurements. Lutz \& Kelker (1973 - LK) undertook the first quantitative
analysis of this effect, which has the same source as Malmquist bias:
if the number of stars is increasing with decreasing true parallax, $\pi_0$, 
then symmetric errors in the measured parallax, $\pi$, will lead a sample
including more stars with overestimated parallaxes, $\pi = \pi_0 + d\pi$, than with 
$\pi = \pi_0 - d\pi$.
That is, $\bar\pi_0 < \bar\pi$; the average distance is underestimated, and, as
a consequence, the average luminosity is underestimated. 
Lutz and Kelker determined specific corrections for
the case of a uniform stellar distribution, i.e. a parallax distribution, 
P$(\pi) \propto \pi^{-4}$.
Smith (1987) has shown that their calculations can be described by the
empirical formula

$$\Delta M_{LK} \qquad = \qquad 5 \times log ( [ 1 +  \sqrt{1 - 19 (\sigma_\pi / \pi)^2} ] / 2)$$

However, Hanson (1979) has demonstrated that the constant-density LK
corrections are seldom appropriate for analysing observational samples,
where magnitude and proper-motion limits can modify the selection effects,
and he derives a more general analytic
representation of the LK corrections. If the parallax distribution can
be characterised as a power law, P$(\pi) \propto \pi^{-n}$, then
the LK correction can be approximated as

$$ \Delta M_{LK} \qquad = \qquad -2.17 \times[(n + {1 \over 2}) {\sigma_\pi \over \pi}^2 \quad + \quad 
({{6n^2 + 10n + 3} \over 4} ){\sigma_\pi \over \pi}^4] $$

(Hanson, 1979: equation 31). The uniform density ($n=4$) approximation
lies slightly below the original Lutz-Kelker datapoints (figure 1). 

If the stellar sample under study is wholly or partially
magnitude-limited, as is the case with the Hipparcos observations of
the Lowell proper motion stars, then one
expects an exponent $n < 4$. As figure 1 shows, a smaller value of $n$ leads to
lower predicted corrections: with fewer stars at smaller parallax, the probability of
overestimating an individual parallax measurement is correspondingly reduced.
The appropriate exponent to use for a given sample can be estimated empirically
using the cumulative proper motion distribution of the sample of stars for
which one has parallax data. If P$(\pi) \propto \pi^{-n}$, 
and the stellar velocity distribution does not vary significantly within the sampling
volume, then N$(\mu) \propto \mu^{-n+1}$.

Hanson (1979) calculates N($\mu$) for the complete Lowell catalogue, and finds 
it consistent with a uniform distribution. However, Hipparcos observations were
effectively limited to stars with $V < 12$ magnitude (80 \% of the
Lowell stars in this magnitude range have parallax measurements).
The cumulative proper motion distribution of the full
Hipparcos sample is N$(\mu) \propto \mu^{-2.4}$ ($n=3.4$). The stars
listed in Table 1 have been selected on the basis that they have parallax 
measurements of higher precision, hence one might expect the inclusion
of a proportionately larger fraction of the nearer stars. This is the case: the
proper motion distribution of the Table 1 stars shows a slower
increase in number with decreasing proper motion, best matched by
N$(\mu) \propto \mu^{-1.2}$ ($n=2.2$). 

The appropriate value of the parameter $n$ to adopt in calculating the
LK absolute magnitude corrections for the Table 1 subdwarf calibrators is
$n=3.4$, since that gives an appropriate characterisation of the 
selection effects which defined the original parallax sample (i.e.
Lowell proper motion stars with V$<$ 11). Were one to analyse the complete
dataset, the Lutz-Kelker effects would be calculated on that basis. 
Selecting stars with high-precision measurements from that parent sample
does not change the underlying systematic bias in the sample.

The resultant Lutz-Kelker corrections amount
to only -0.13 magnitudes for a star whose parallax has an uncertainty of 
12 \%, but rise to -0.43 magnitudes as the precision drops to 20 \%. In
comparison, the $n=2$ calibration (applied by S96 to their subdwarf sample)
leads to corresponding corrections of -0.08 and -0.26 magnitudes - a substantial 
difference for the lower precision measurements. On this basis, we have divided
our sample into primary calibrators with ${\sigma_\pi \over \pi} < 12 \%$, which 
we use to estimate directly the cluster distances, and the lower precision
secondary calibrators, which serve to check
the derived distance moduli. Averaging over the complete
sample of 15 Hipparcos and 3 YPC stars listed in Table 1, there is a
systematic offset of only 0.02 magnitudes between the $n=2$ and $n=3.4$ 
$\Delta M_{LK}$ calibrations. Since we weight each star by its uncertainty in
matching the fiducial cluster sequences (section 3), the derived
distance moduli are affected by no more 0.01 magnitudes.
It is clear from figure 1, however,  that the choice of the correct
exponent is of considerably greater significance if one is faced with analysing
a sample including many stars with parallaxes of lower precision.

\subsection { A comparison with previous ground-based parallax measurements}

The subdwarfs that are employed in cluster main-sequence fitting
were selected originally on the basis that their parallaxes were
of the highest precision available. Thus, as with astronomical site testing,
there is likely to be a bias towards over-optimism, that is, stars where the 
measured parallax, $\pi$, and hence $\sigma_\pi \over \pi$, is overestimated. 
Therefore, as more parallax
data of higher accuracy are accumulated, one expects $\pi$ to decrease
as it converges towards $\pi_0$, and the inferred absolute magnitude to 
become brighter. A comparison of the various absolute magnitudes cited in
the astronomical literature for the ``classical"
subdwarfs, such as HD 201891 and HD 1394439/40, shows that this is indeed the
case, and with Hipparcos providing up to a factor of ten increased accuracy, it
is not surprising that the trend continues. 

Figure 2a compares the new Hipparcos data against the absolute parallax measurements
listed for 26 stars in the YPC. The formal uncertainties in the YPC measurements
range from 1.5 to 15 milliarcseconds (mas), and the mean difference between the two
datasets is -0.60 $\pm$ 5.09 mas. The trend towards negative residuals at smaller
$\pi_{abs}$(YPC) reflects our selection of only the higher-precision parallaxes:
the Hipparcos parallaxes for stars with $\pi$(YPC) $> \pi$(Hip) and $\pi$(YPC) $<$ 10 mas 
are less precise than 20\% cutoff we have imposed on our sample. 
Note, however, the tendency towards
positive residuals amongst the stars with larger YPC parallaxes. In nine out of the eleven
cases where the YPC parallax exceeds 15 mas, the Hipparcos parallaxes are smaller. 

This offset towards smaller parallaxes and larger inferred distances
can be explained to a large extent by
the higher accuracy of the Hipparcos data, and the consequent reduction in
the necessary systematic (Lutz-Kelker) parallax corrections that are required
for statistical analysis. However, given the small number of calibrators available
to previous studies, significant offsets in the mean M$_V$ calibration can
persist even after allowing for LK effects. As an example, figure 2b compares the 
ground-based and Hipparcos LK-corrected absolute magnitudes for nine subdwarf
included as calibrators by S96 in their study of M5. Eight of these stars have
high-precision Hipparcos parallax measurements, and we have also plotted data for 
HD 219617, although the substantial
uncertainty in even the Hipparcos parallax leads to its exclusion from any further
statistical analysis. 
The mean offset in absolute magnitude between the two
datasets for the eight high-precision subdwarfs is 0.094$\pm$0.094, in the sense
that the Hipparcos stars are brighter. 
Including HD 219617, the mean magnitude difference, $\Delta M_V$, becomes
0.159$\pm$0.105 magnitudes. We also compare in figure 2b the S96 and 
YPC LK-corrected absolute magnitudes for HD 25329, 134439 and 134440. While there
is little difference in the case of HD 25329, the final
YPC data lead to absolute magnitudes for the last two stars which are 0.1
magnitude brighter than those adopted by S96.

The mean residuals between the S96 and Hipparcos-based absolute magnitudes are
formally consistent at the 1$\sigma$ level. However, note that S96 based
their selection of the Hanson $n=2$ LK corrections on the proper-motion
distribution of their subdwarf calibrators. As described in the previous
section, the bias towards higher-precision measurements amongst those
stars could lead to adopting too small a value of n. If we recalculate
M$^{LK}_V$ for the S96 stars using $n=3$, the mean difference is only
0.04 magnitudes; for $n=3.4$, the difference is +0.02 magnitudes. 
In any case, the presence of this systematic difference
between the (significantly more accurate) Hipparcos measurements and the S96
parallax dataset has obvious repercussions for determining the distances of
globular clusters.

\subsection {The metal-poor main-sequence}

Most of the subdwarfs in the current sample have accurate photometric
observation in the UBV passbands. This is unfortunate, since the relatively
steep slope of the (M$_V$, (B-V)) main-sequence and the sensitivity
to line-of-sight absorption are significant factors in estimating distances to
many clusters, as is discussed further in the following section. However, we can
at least use our data to provide an empirical definition of the position of the
upper main-sequence as a function of abundance. 

Figure 3 plots the (M$_V$, (B-V)$_0$) colour-magnitude diagram defined by
the stars in our sample with high-precision ($\sigma_\pi \over \pi$ $< 12 \%$)
trigonometric parallaxes. We have used different symbols to identify stars
within different abundance ranges, and have also plotted data for subdwarfs
with -1.0 $\le$ [Fe/H] $\le -1.25$ and for stars within 30 parsecs of the Sun
which are included in our subset of the Hipparcos catalogue. The most significant
feature of this diagram is the small separation in M$_V$ between the [Fe/H]$\sim -1.75$
and [Fe/H] $< -2$ subdwarfs. As discussed further in sections 3.3 and 4.2, where
we compare our data against model isochrones, this suggests that current models 
overestimate the change in luminosity with decreasing abundance for extreme
metal-poor subdwarfs. [Fe/H]=-2.0 subdwarfs are not as subluminous as
expected. Again, this has clear implications for the cluster distance scale.

\section {Globular cluster distance determinations}

The high-precision parallax subdwarfs listed in Table 1 span a range of almost 1 
dex in [Fe/H] and absolute magnitudes from M$_V$=+4 to +7, defining the upper
main-sequence for intermediate and extreme low-abundance systems. We can supplement
these observations with data for three other subdwarfs with accurate ground-based 
parallax measurements: the intermediate subdwarf binary HD 134439/134440
($\sigma_\pi \over \pi$ = 0.05); and the extreme subdwarf HD 25329 ($\sigma_\pi 
\over \pi$ = 0.03). We have taken the photometry and abundance
estimates for the latter three stars from S96, while the parallaxes are from the YPC. 

\subsection{Main-sequence fitting}

We have matched the local calibrators against fiducial colour-magnitude sequences 
determined for seven globular clusters - the intermediate abundance systems M5 (S96), 
M13 (Richer \& Fahlman, 1986; Sandage, 1970 (subgiant and RGB)) and NGC 6752
(Penny \& Dickens, 1985); 
and the metal-poor systems M30 (Richer, Fahlman \& Vandenberg, 1988), 
M92 (Stetson \& Harris, 1988),
M15 (Durrell \& Harris, 1993; Fahlman, Richer \& Vandenberg, 1985) and 
M68 (McClure et al, 1987). All have (B, V) CCD photometric observations which extend
well below the main-sequence turnoff. (Most, unfortunately, have {\sl only} 
published BV observations.)
While all of these clusters lie at moderate to high galactic latitude, at least three clusters
have significant foreground reddening. Determining the magnitude of foreground reddening
is not straightforward, and the value adopted
can have a significant impact on one's conclusions, as
will become apparent. Two widely used estimators, which have the advantage of
being available for most clusters, are Burstein \& Heiles (1982) HI maps and
Zinn's (1980) analysis of reddening-free colour indices derived from integrated photometry. 
The uncertainties associated with the individual measurements on either scale are $\pm 0.02$
to 0.03 mag, so the two sets of results are in statistical agreement. However, a 
difference of 0.05 mag in E$_{B-V}$, as is the case for M68, has a substantial
effect on the derived intrinsic luminosity of the horizontal branch and the colour at 
the turnoff, and hence the inferred age. 

Of the seven clusters in our sample, there is general agreement that M5, M13 and M92 have 
little foreground absorption, but there is less than unanimity on the appropriate
reddening for the other four clusters. Table 2 lists other reddening estimates for
those systems. We have not included any estimate based on inter-cluster
comparisons, since those are generally predicated on the assumption that 
clusters of the same abundance have identical colours at the turnoff (i.e. identical 
ages). This technique obviously hampers the prospect of detecting age differences
between clusters. Thus, we have concentrated on results based on the photometric
properties of horizontal branch stars or nearby field stars. As with the Burstein \& Heiles
and Zinn results, the derived reddenings span a range of $\sim$0.04 magnitudes for
an individual cluster, even if based on the same method. Thus, Brocato et al (1994)
estimate a reddening of E$_{B-V} = 0.03$ mag towards M68 (and 0.07 towards M15) based
partly on the colour of the blue edge of the RR Lyrae instability strip; Walker (1994),
who has undertaken the most thorough examination of this issue, estimates E$_{B-V} = 0.08$ 
mag from the same parameter (derived directly from his photometry). Walker's final estimate
of E$_{B-V} = 0.07 \pm 0.01$ mag for M68 is clearly better supported observationally than is
the lower value favoured by Brocato et al. 

Clearly, deep I-band and infrared
(JHK) photometry would serve to minimise these problems. However, lacking such data, we
have matched the cluster sequences to the local calibrators for a range of assumed
values for the foreground absorption, listed in Table 3.
In particular, for the metal-poor clusters, the
three values adopted for E$_{B-V}$ represent the maximum and minimum likely reddening,
and a median value which results in a colour-magnitude
diagram that is in close agreement with the de-reddened M92 fiducial sequence.
The three low-reddening clusters, M5, M13 and M92, serve 
as our prime references for comparisons with theoretical isochrones and for age determinations.

Table 3 also gives the Zinn \& West (1984 - ZW84) abundance estimates for each system. 
Recently, Carretta \& Gratton (1997) have published abundances for 24 clusters,
including all five in our sample, based on high-dispersion spectroscopy. Their
results are within 0.1 dex of the ZW84 scale for the metal-poor clusters, but
indicate metallicities that are $\sim 0.2$ dex higher than the ZW84 values for
M5, M13 and NGC 6752. Were we to adopt these higher abundances for the latter clusters,
then we should also calibrate their distances using higher-abundance subdwarfs,
and would expect to derive distance moduli that are larger by 0.05 to 0.1 magnitudes.
However, the issue is complicated further by the fact that Balachandran and
Carney (1996) have used echelle spectra to derive an abundance of -1.22 for HD 103095 
- that is, 0.2 dex higher than the CLLA analysis. Given that most
of the calibrating subdwarfs in Table 1 lack high-dispersion spectroscopy, we 
have adopted the ZW84 and CLLA abundance scales respectively for the clusters
and subdwarfs in the current analysis.

We have adjusted each of the observed (V, B-V) fiducial sequences to intrinsic
colours and magnitudes using the reddenings listed in Table 3, adopting a
value of 3.1 for the ratio of total to selective extinction. The
best-fit distance modulus is determined by (linearly) minimising the residuals in
(V$_0$-M$_V$) with respect to
the relevant subset of primary calibrators. In the case of M5, the latter stars
span the abundance range -1.25 $<$ [Fe/H] $<$ -1.6; for M13, we adopt abundance limits
of -1.4 $<$ [Fe/H] $<$ -1.85; and we include all stars with [Fe/H] $>$ -1.7 for
the three extreme metal-poor clusters. In the fitting process, each star is
weighted by the uncertainty in its absolute magnitude - that is, 
the residuals are normalised based on the individual values of $\sigma_M$.

In this initial analysis, we have not applied the usual colour-correction technique to 
adjust all of the stars to the same effective
abundance, since the small abundance range in each sub-sample means that the maximum
correction is less than 0.015 magnitudes in (B-V). We have, however, estimated the 
effect of errors in the subdwarf colours by applying systematic offsets of 
$\pm 0.02$ magnitudes in (B-V) to {\sl all} of the calibrators in each group.
As one might expect given the steep slope of the (M$_V$, (B-V)) main-sequence,
these offsets lead to a typical change in distance modulus of $\sim 0.14$ magnitudes,
in the sense that making the subdwarfs bluer reduces the inferred distance.
Changing the reddening by $\pm 0.02$ mag generally leads to somewhat smaller changes in
distance modulus ($\pm 0.1$ mag), since the change in V$_0$ counteracts to some extent
the shift in colour, $\delta E_{B-V}$, of the de-reddened cluster sequence. A lower
value for the reddening leads to a smaller distance modulus, and an inferred
fainter intrinsic luminosity at the cluster turnoff. The change in distance modulus 
also depends on the distribution of calibrating subdwarfs, however, and, with
more higher-luminosity calibrators (at colours where the main-sequence is steepening),
distance determinations to the extreme metal-poor clusters are more sensitive to
changes in the foreground reddening.

Figures 4 and 5 show the results of the main-sequence fitting. In each case, we 
plot the calibrating subdwarf data together with the individual cluster
colour-magnitude diagrams offset by the best-fit distance modulus, (m-M)$_{12}$
(Table 3). The upper panel in figure 4 plots the calibrated NGC 6752 colour-magnitude
diagram for the two estimates of the reddening given in Table 3, while
the separate panels in figure 5 show results of adopting either the 
mean or maximum reddening values for the metal-poor systems. Clearly, the
distance moduli that we derive for the latter clusters are dependent sensitively
on the adopted extinction, a matter we discuss further in 
the following section while comparing our new results to older analyses.
However, note that adopting the minimum possible value of E$_{B-V}$=0 for
M92 reduces the derived distance modulus by only 0.07 magnitudes, to 
14.86 magnitudes. 

\subsection {A comparison with previous distance determinations}

Table 3 presents distance determinations derived from previous studies of
the clusters in our sample.
In the case of the intermediate abundance systems, our new results differ
by no more than $\sim 0.15$ magnitudes from currently-accepted values. This
is clearly comparable to the offset between the S96 and Hipparcos-based
absolute magnitudes, as shown in figure 2. However,
one should note that the S96 analysis of M5 is based on the high-dispersion
abundance estimate of [Fe/H]=-1.19. Dropping the abundance to [Fe/H]$\sim$-1.4,
as in our analysis, reduces their derived distance modulus by $\sim0.1$ magnitude
and increases the disparity with respect to our result. As regards M13 and NGC 6752,
two clusters with similar abundances, we note that aligning the fiducial
sequences in colour at the turnoff (i.e. E$_{B-V} = 0.02$ mag for NGC 6752) leads
to an offset of $\sim 0.15$ magnitudes on the subgiant branch 
(NGC 6752 fainter) and an $\sim 0.05$
magnitude difference in the colour of the giant branch. A direct comparative photometric
study of these two clusters would clearly help settle whether the morphological
differences in the shape of the turnoff region 
reflect intrinsic differences or subtle problems with instrumental calibration.

Reddening is clearly an important factor in the analysis of several of the 
metal-poor clusters (figure 5). 
Nonetheless, all of our derived distances are larger than previous determinations.
Given these differences, we have tested the robustness of our conclusions by
re-analysing the [Fe/H]$<$-2.0 cluster data (particularly M92), modifying both the set of stars 
in the calibrating sample and the distance-determining technique. First, we have
limited the subdwarf calibrators 
to only those stars with [Fe/H] $< -2$, thereby eliminating the influence of the
three slightly more metal-rich subdwarfs at (B-V)$\sim 0.6$. The best-fit
distance moduli are reduced, but by only $\sim$0.03 magnitudes. Indeed, matching
M92 against either subdwarfs in the abundance range
-1.7 $<$ [Fe/H] $<$ -1.95 gives (m-M)$_0$=14.92,
further emphasising our comments of section 2.3 on the apparent "saturation"
of subluminosity with decreasing abundance.

Second, eleven stars (four primary and seven secondary calibrators) listed in Table 1 
are identified by CLLA as suspected or confirmed single-lined binaries based
on their radial velocity measurements, while analysis of the Hipparcos
astrometry leads to three other stars being suspected as non-single. \footnote
{Twelve potential single-lined binaries amongst 35 halo subdwarfs is a
surprisingly high fraction, given that CLLA cite an overall spectroscopic-binary
frequency of only 15 \% for metal-poor stars.} As single-lined stars, several
may have companions that are substantially less massive, but one of the
suspected binary stars is
HD 84937 (M$_V$=3.73), which appears over-luminous for its
colour when compared against the cluster sequences. However, excluding all possible
binaries from the main-sequence fitting affects the derived distance moduli
at only the $\pm 0.05$ magnitude level. 

Third, cluster main-sequence fitting analyses usually exclude subdwarf calibrators
with M$_V < 5.5$ on the basis that the position in the HR diagram of these
stars is dependent on their age. The evolutionary variation in (B-V) colour is less
than 0.02 magnitudes for M$_V > 4.5$ and ages of less than 10 Gyrs, however.
Furthermore, 
if the subdwarfs are younger than the clusters, then they lie blueward of the
correct isochrone and will lead to an underestimate of the distance and
an overestimate of the age. Given the many other uncertainties in the main-sequence
fitting process, there is little justification for excluding stars such as HD 19445 
from the analysis on the basis of possible evolutionary effects. (A more valid reason
is the steepness of the main-sequence at these magnitudes, and consequent sensitivity to
inaccurate photometry.) In any event, excluding
all metal-poor stars with M$_V < 5.5$ (whose position in the H-R diagram is dependent
significantly on age) from the calibrating sample still leads to a best-fit distance
modulus of 14.9 magnitudes for M92. 

Fourth, we can take the opposite approach, and use only HD 19445 as a distance
estimator. The abundance is a close match to the metal-poor clusters; it has
a well-defined absolute magnitude of M$_V = 5.12 \pm 0.09$; there is
no evidence that HD 19445 is an unresolved binary star; and, as we noted 
above, if HD 19445 will only lead to an overestimate of the cluster
distance if it is older (redder) than the cluster stars. Hence, this
star can be taken as representative of an unevolved, extreme F-type subdwarf.
The M92 fiducial sequence crosses (B-V)=0.477 ( (B-V)$_0$=0.457) at V=20.0. 
Comparison with HD 19445 alone
then gives a distance modulus estimate of $\sim 14.80$ - substantially higher
than the 14.65 derived by Bolte \& Hogan (1995). Thus, the
metal-poor stars in the Hipparcos sample define an empirical main-sequence which 
is well-matched to the M92-like cluster main-sequence at all colours. 

As an alternative to trimming the number of calibrators, we have
repeated the main-sequence fitting using the complete sample of subdwarfs
listed in Table 1. We have used the same abundance limits as in the
analysis based on the primary calibrators, again using the uncorrected
(B-V) colours for the subdwarfs. The derived distance
moduli are listed in Table 2 as (m-M)$_{20}$. All indicate higher
distances than the analysis limited to the high-precision stars. 

A systematic difference between our photometric data and the magnitudes and colours
used in previous analyses would produce a systematic difference in the estimated distance.
Our (B-V) colours and V magnitudes are taken from CLLA. While there is
excellent agreement between the magnitudes in the latter catalogue and the data 
listed both in the Hipparcos catalogue and by S96, the (B-V) colours for a few stars 
differ by 0.02 to 0.03 magnitudes. However, in almost every case, the difference is in
the sense that the CLLA colours are {\sl bluer}. Thus, if we were to adopt the
alternative colour estimates, the distance moduli would be driven to larger
values, by from 0.1 to 0.15 magnitudes. 

As a final test, we have used theoretically-based colour correction to
derive mono-metallicity subdwarf sequences, and matched these against the
cluster data. First, we have limited our sample to primary Hipparcos
calibrators with [Fe/H] $< -1.35$ (i.e. all save HD 194598). 
Following S96 and BH95, we have used the BV92 theoretical isochrones to compute
the differential colour corrections required to place all of those stars 
on an [Fe/H]=-2.1 main sequence. The required corrections, $\delta_{B-V}$, are listed in
Table 1. We then used the same minimisation technique to match the (V$_0$, (B-V)$_0$)
fiducial sequences of the metal-poor clusters to the mono-metallicity subdwarf distribution.
The resulting distance moduli, (m-M)$_{-2.1}$, are listed in Table 2, and
figure 6 compares the calibrated sequences and the calibrators.
While this fitting technique does produce lower distance estimates, the reduction in
(m-M)$_0$ is only $\sim$0.1 magnitude. 

We have also undertaken the same analysis using a subset of the S96 
subdwarfs as calibrators. Rather than basing the calibration on
stars with the highest-precision parallax measurements, we have 
selected a sample of metal-poor stars with parallaxes of at least
moderate accuracy. Limiting the calibrating sample to the more metal-poor
stars limits the required colour corrections. Figure 1 shows that the
difference between the $n=2$ and $n=3$ Hanson formulations is only
0.06 magnitudes at $\sigma_\pi \over \pi$=15\%. Thus, systematic errors
in the Lutz-Kelker corrections are not a significant factor for stars
with parallaxes of this precision. S96 include
eleven subdwarfs in their compilation (including HD 103095) with parallaxes 
satisfying that criterion and with [Fe/H]$<$-1.3. We have used these 
stars to calibrate the distance 
to M92, taking the relevant data directly from Table 9 of S96.
We have adopted both their (B-V) colours, 
adjusting each to [Fe/H]=-2.1 using the BV92 isochrones, and the 
LK-corrected absolute magnitudes listed in that paper (i.e. for a Hanson
n=2 model). We have not adjusted the parallax measurements of HD 25329, 134439
and 134440 to the YPC values, but applying our fitting technique, we derive
a best-fit true distance modulus of 14.82 for M92 (figure 7). D'Antona et al (1997)
have derived similar results based on comparable pre-Hipparcos, deriving a
a distance modulus of 14.80 magnitudes for M92. As they note, HD 103095 falls 
$\sim 0.2$ magnitudes below the main-sequence if one minimises the residuals 
(V$_0$ - M$_V$) with respect to {\sl all} of the subdwarfs in the sample. 

Thus, all of the analyses based on the new Hipparcos parallax data, and even
analyses based on the average of the older ground-based data,
indicate that the current estimates of the distances of intermediate-abundance 
([Fe/H]$\sim -1.5$) clusters may require adjustments of up to 6 \%, 
while those to the metal-poor clusters M92, M15, M30 and M68
fall short of the mark by 15 to 20 \%.

\subsection{Empirical vs. semi-empirical distance estimates}

Why are our current distance modulus estimates so much larger than most
previous estimates? There are three main reasons: the 
higher accuracy and precision of the new observations; the larger range in M$_V$ 
spanned by the new data; and, most important in the case of the extreme
metal-poor clusters, the increased number of [Fe/H]$<$-1.7 calibrators,
eliminating the necessity for theoretically-based colour corrections. 

In most previous studies, the distance modulus derived from main-sequence fitting
rests on only 7 to 10 calibrating parallax stars. For example, Bolte \& 
Hogan (1995) relied on
ten nearby subdwarfs in calibrating the distance to M92. Seven of these stars
have [Fe/H] $>$ -1.5, and only two of the ten are more luminous than M$_V$=6. 
The S96 compilation of ground-based parallax data represented a significant 
improvement over previous studies, but even this sample of 23 stars includes 
only seven with parallax uncertainties of better than 7 \%. In contrast, 
eight stars have ${\sigma_\pi \over \pi} > 20 \%$, including three of
only seven stars with [Fe/H]$<$-2. As figure 1 shows, these last stars lie in a r\'egime where
the predicted LK corrections are dependent strongly on the functional form assumed for the
parallax distribution of the parent sample, and the inferred absolute magnitudes are correspondingly
uncertain. Finally, four of the high-precision stars have [Fe/H]$>$-1.6. 
Hence, in matching a metal-poor cluster, one is dependent on the accuracy of the 
colour corrections applied to the latter stars which, as the most accurate, 
dominate any least-squares fitting process. 

To demonstrate the problems faced by previous studies, figure 8 plots the (M$_V$, (B-V))
colour-magnitude diagram described by the ten calibrators used by Bolte \& Hogan (1995)
in their study of M92. We have indicated the abundance of each star and plotted
the [Fe/H]=-1.48 and -2.26 12, 14 and 16 Gyr. isochrones from the BV92 models.
As discussed further below, the latter have an enhanced oxygen abundance (of 
[O/Fe]=+0.6 at [Fe/H]=-1.48 and +0.75 at -2.26), but do not allow for any enhancement of 
the $\alpha$-elements (Si, Ca, Ti etc).
However, apart from the most luminous star, HD 201891, 
there appears to be good agreement between the [Fe/H]$\sim$-1.4 subdwarfs (which
dominate the sample) and the [Fe/H]=-1.48 isochrone. Given this agreement, it is not
surprising that, after computing colour corrections from the BV92 models, 
the mono-metallicity [Fe/H]=-2.26 subdwarf sequence (open circles
in figure 8) is a good match to the [Fe/H]=-2.26 BV92 isochrone. 
Bolte \& Hogan adopted the semi-empirical approach of 
adjusting the BV92 isochrones by +0.015 in (B-V) to fit
HD 103095, the star with the most accurate parallax measurement, and then
determined the distance modulus to M92 ((m-M)$_0$=14.65) by fitting the de-reddened cluster
sequence to the isochrone data. 

Thus, Bolte \& Hogan's result is crucially dependent both on the accuracy to
which the BV92 isochrones match subdwarfs of the same abundance, and on the accuracy
to which HD 103095 ([Fe/H]=-1.44) can be transformed to match the (B-V) colours of
an [Fe/H]=-2.26 subdwarf. The Hipparcos data allow us to test these hypotheses
by comparing the BV92 (M$_V$, (B-V)) isochrones directly against the distribution of
the local subdwarfs. Figure 9 shows this comparison for isochrones of four specific
abundances. In each case, we have included the 12, 14 and 16 Gyr isochrones presented
by BV92 without any adjustment to the (B-V) colours, and plot both primary and
secondary subdwarf calibrators (from Table 1) of similar abundance. We have also
plotted the colour-magnitude diagram for the appropriate low-reddening cluster at
(m-M)$_{12}$. It is clear that the BV92 isochrones are significantly bluer than
the subdwarf data at all abundances. Specifically, the [Fe/H]=-2.26 isochrones
are $\sim 0.06$ magnitudes bluer than the calibrated M92 sequence at M$_V$=6.5, the
absolute magnitude of HD 103095. Thus, adjusting these isochrones by only
+0.015 leads to an offset of $\Delta M_V \sim +0.2$ magnitudes, and a correspondingly
smaller distance modulus when one matches the isochrones against the M92 data. Since
the isochrones are too blue for the unevolved stars, matching the colour at the 
turnoff inevitably leads to age estimates of more than 14 Gyrs. 

Increasing the [$\alpha$/Fe] ratio in these models mitigates the offset in 
colour to some extent, since an
enhanced $\alpha$-element abundance moves a model of given age and
abundance to redder colours and fainter magnitudes. A full set of 
models remains to be calculated, but preliminary calculations by Vandenberg (reported by S96) 
show a systematic offset of up to $\sim 0.03$ magnitudes in (B-V) for 
intermediate-abundance lower main-sequence stars (S96, figure 16). The
[Fe/H]=-1.31 $\alpha$-enhanced isochrones lie close to the oxygen-enhanced
[Fe/H]-1.03 BV92 tracks. S96 used Vandenberg's calculations to determine a distance modulus of
(m-M)$_0$ = 14.41 for M5 (and an age of 12 Gyrs). This is significantly
closer to the result that we derive by direct comparison against the local
subdwarf stars. 

The comparison plotted in figure 9 clearly invalidates the direct use of the
BV92 isochrones to determine cluster distance moduli (as in Durrell and Harris, 1993). 
However, the subdwarf data also suggest that these models may not be
well suited for the calculation of the differential colour corrections required to
adjust stars to an extremely metal-poor mono-metallicity sequence. As already
noted in section 2.3, there is relatively little change in $\delta M_V$, the
distance that a star falls below the solar-abundance main sequence, for
[Fe/H] $<$ -1.5 dex. As we have already noted, the best-fit M92 distance
modulus is essentially identical whether one uses the most metal-poor stars
or -1.7 $<$ [Fe/H] $<$ -1.95 subdwarfs as calibrators. 
One might attribute this to problems in the abundances
derived for the metal-poor subdwarfs that have been added to the "classical"
calibrators. However, HD 103095, HD 64090 and HD 25329 give individual distance
estimates to the intermediate-abundance cluster M13 which agree to 
within 0.05 magnitudes despite the difference in abundance amongst the subdwarfs.
Indeed, even using only HD 19445 ([Fe/H]=-2.1) as a calibrator
changes the inferred distance modulus to M13 by -0.15 magnitudes. This is
less than the $\sim 0.25 - 0.3$ magnitudes one would expect based on 
the theoretical isochrones and, again, suggests that the BV92 isochrones 
may overestimate the variation in luminosity and colour at the lowest abundances.

\subsection {Summary }

It is clear that the currently-accepted distances to the metal-poor clusters
are tied closely to the assumption that standard stellar models 
(particularly the BV92 dataset) 
provide an accurate representation of the colour-magnitude distribution of
metal-poor stars. The Hipparcos data indicate that this is not the case.
Empirical distance determinations, even based on a suitably chosen sample drawn from
the S96 calibrators, lead to higher distance moduli for most the globular
clusters in our sample. It is also clear that foreground reddening introduces
significant uncertainties in the distance determination, particularly for the
more metal-poor systems.

Taking these considerations into account, we have adopted the
(m-M)$_{12}$ distance moduli listed in Table 3 as our current
best estimate of the distances of the seven clusters in the current sample. 
Of the three distance estimates given for M15, M68 and M30, we regard
the low-reddening value as the least likely, and take the median
value as our reference point. 
The continued scarcity of extremely metal-poor stars with high-precision
parallaxes means that the distances to the [Fe/H]$\sim$-2.1 clusters are
least well determined. Tying the distances to HD 19445, rather than
to the average of the extreme metal-poor calibrators, decreases the distance
moduli by 0.12 magnitudes. We regard this as an upper limit to the likely
systematic errors in the distances to these clusters.

Summarising, we find M5 and M13 to lie at respective distances of 7.8 and 7.9 kpc (distance
moduli of 14.45 and 14.48), and the distance to NGC 6752 is estimated as 4.3 kpc.
Adopting the median reddening estimates for M68, M15 and M30, we place these
clusters at distances of 11.4, 11.9 and 9.8 kpc ((m-M)$_0$ = 15.29, 15.38
and 14.95), while M92 is estimated to be at a distance of 9.7 kpc
(distance modulus of 14.93 mag). Our estimated random uncertainties in
these values are $\pm 0.1$ mag in (m-M)$_{12}$, or $\pm 5 \%$ in distance.

\section {Discussion}

Based on the analysis presented in this paper, we have arrived at the conclusion
that the metal-poor globular clusters lie at larger distances than was
accepted previously. The discrepancies are particularly pronounced for
the extreme metal-poor systems, where our estimates exceed the standard value
by $\sim 15\%$. This conclusion obviously affects the luminosities that one
derives for stars in these clusters, specifically on the horizontal branch
and at the main-sequence turnoff, and adjusting the latter clearly affects the 
age that one infers for these systems. Our revised distances also suggest that the globular
clusters are younger than the currently-accepted value of $\sim$16 Gyrs. 
Feast \& Catchpole (1997) have arrived recently at 
similar conclusions, but through a complementary series of observations. They have
used their proposed revision of the distance to the LMC to infer the absolute
magnitude calibration of RR Lyrae variables, and thence re-determine distances to 
globular clusters. Our analysis runs in the opposite direction: we use our
distance estimates to the globular clusters to infer the RR Lyrae absolute
magnitude calibration, and the distance modulus of the LMC. In this
section we explore some of the immediate repercussions of this re-calibration of the
cluster distance scale. 

\subsection{The absolute magnitude of RR Lyraes and the distance to the LMC}

Moving the globular clusters to larger distances results in an increased luminosity
for the horizontal branch stars, and implies a change in the RR Lyrae (M$_V$, [Fe/H])
calibration. The horizontal branch in M13 lies predominantly blueward of the instability 
strip, and few variables have been detected. However, four of the remaining clusters
all have populous, well-studied RR Lyrae populations. 

Considering first the metal-poor clusters, 
Carney et al (1992b) have presented BV photometry of seven RR Lyraes in
M92, and the resultant mean (intensity-weighted) magnitude is $<V>= 15.103 \pm 0.027$.
Allowing for a reddening of E$_{B-V}$ = 0.02 mag., and adopting a true
distance modulus of 14.93 magnitudes, we derive 
a visual absolute magnitude of M$_V$ = 0.11 magnitudes. 
Walker (1994) derives a mean apparent magnitude of 15.64 magnitudes from
observations of 40 stars in M68. 
Combined with the mean value for the line-of-sight extinction of
A$_V$= 0.15 mag and (m-M)$_0$=15.29, the inferred absolute magnitude
is M$_V$=0.20 magnitudes. The RR Lyraes in the
third metal-poor cluster, M15, have been surveyed most recently by Silbermann \&
Smith (1995), and their observations give a mean magnitude of $<V> = 15.83 \pm 0.09$ mag 
for 28 stars with well-defined light curves, implying M$_V$=0.18, for the mean
extinction solution (E$_{B-V}$ = 0.09 mag) listed in table 3.

The extensive M5 RR Lyrae population has been studied in detail most
recently by Reid (1996), who used CCD observations to map the light curves
of over 50 stars. Those data give
(intensity-weighted) mean apparent magnitudes of 
$<V> = 15.059 \pm 0.064$ for 33 RRab variables  and $<V> = 15.039 \pm 0.037$ for 
11 RRc variables. (These averages exclude five higher-luminosity variables which
may be on the return traversal of the horizontal branch.) S96 derive similar
mean magnitudes from their less extensive CCD observations of a smaller sample
of variables. Combined with a reddening of 
0.03 magnitudes and (m-M)$_0$ = 14.45 mag., the latter estimates give
gives M$_V$ = 0.51 magnitudes. 

Figure 10 compares these two absolute magnitude estimates against several
recent proposed (M$_V$, [Fe/H]) calibrations. Our results 
imply a luminosity-abundance dependence in
the absolute visual magnitudes of RR Lyrae variables which is steeper even than that derived
by Sandage (1993), although we would emphasise that there is no compelling evidence
that this relation should be extrapolated linearly into the metal-rich ([Fe/H] $>$-1) 
domaine. If the correlation were to flatten significantly for [Fe/H]$>$-1.6, then
the current results may well be consistent with the Layden et al statistical
parallax analyses. Two-thirds of the local halo RR Lyraes fall in this
intermediate abundance range, and have a correspondingly large influence on the
mean absolute magnitude derived for the sample as a whole.

In any case, if we assume that the RR Lyraes in metal-poor LMC clusters are
similar to the Galactic variables, we can use our results to estimate a
distance to the LMC. Walker (1992) lists intensity-weighted average magnitudes for
variable in five LMC clusters with [Fe/H]=-1.8$\pm$0.1. His analysis is based on
only nine star for NGC 2210, but Reid \& Freedman (1994) provide average magnitudes for
over 30 cluster members, and combing these datasets gives the mean magnitude of
$<V_0> = 18.98$ cited in the introduction. 
Interpolating in metallicity between the two datapoints 
defined by our observations - M$_V \sim 0.1 \pm 0.1$ at [Fe/H] $\sim -2.1$ and 
M$_V \sim 0.51\pm 0.1$ at [Fe/H] $\sim -1.4$
leads to an absolute magnitude estimate of M$_V \sim 0.3 \pm 0.1$ for the LMC RR Lyraes.
This implies a distance modulus of 18.68 mag. Of the two extremely metal-poor clusters
surveyed by Walker, NGC 1786 ([Fe/H]=-2.3) has a relatively small RR Lyrae population,
and the mean magnitude, $<V_0 = 19.05$ mag, is based on data for only nine stars. The
instability strip is better defined in NGC 1841 ([Fe/H]=-2.2), where observations of 22 stars
give $<V_0> = 18.75$ mag, or, comparing the variables directly with the Galactic
clusters, (m-M)$_0$ = 18.6 mag. Weighting these results leads to a mean LMC distance modulus
of 18.65 mag, ($r \sim 53.7$ kpc), clearly in good agreement 
with the Cepheid-based distance scale, particularly given the recent revision to the
latter proposed by Feast and Catchpole (1997). 

\subsection {Globular cluster ages and galaxy formation }

\subsubsection { Isochrone-based age estimates}

The larger distances that we have derived for most of the clusters in our
sample inevitably lead to higher luminosities for stars at the main-sequence
turnoff and, as a result, implied higher masses and younger ages. The effect is
particularly significant for the metal-poor clusters, where our revised distance
moduli are higher than currently-accepted values by $\sim 0.3$ magnitudes. A
comparison with the isochrones predicted by theoretical modelling offers the
only method of quantifying these differences. The individual stellar model calculations,
however, generate luminosities and temperatures as a function of age. These must be
transformed to the colour-magnitude plane by applying bolometric corrections and
appropriate temperature-colour calibrations before the models can be compared with
observations. The (T$_{eff}$, colour) relations remain somewhat problematical for
metal-poor stars, and small inaccuracies in the transformed colours can have a 
profound impact on the inferred cluster ages, as is clearly demonstrated by
the results plotted in figure 9.

Several recent globular cluster studies (notably BH95, S96) have been based 
on the Bergbusch and Vandenberg (1992) oxygen-enhanced  theoretical isochrones.
These calculations have been criticised, however, since the
[O/Fe] ratio in the models increases with decreasing abundance, from +0.5 dex at [Fe/H]=-1
to +0.75 dex at [Fe/H]=-2.26. The (T$_{eff}$, colour) transformation is based on
Vandenberg's (1992) formalism, and we have already shown that the isochrones do not match
the (M$_V$, (B-V)) distribution of the local subdwarfs. Given this mismatch, we
have considered isochrones generated by two other sets of model calculations.

Straniero \& Chieffi (1991 - SC91) have calculated a suite of isochrones
for abundances between [Fe/H] = -0.5 and [Fe/H] = -2.3, and for ages of 10 to 20
Gyrs. Their models are "classical" in the sense that they allow for enhanced
abundances of neither oxygen nor any of the other $\alpha$ elements. However,
both SC91 and DCM point out that, for modelling purposes, the increased abundances 
of this subset of elements may simply mimic a lesser increase in the overall [Fe/H].
We have allowed for this in matching the models to the cluster colour-magnitude data. 
The transformation of the SC91 (L, T$_{eff}$) data to the observational plane
relies on relations derived by Vandenberg \& Bell (1985) and by Bell \& Gustafsson (1978),
with the latter adjusted to match the former in the region of overlap. Figure 11
compares the resultant (M$_V$, (B-V)) isochrones for ages of 10, 12, 14 and 16 Gyrs
against the Hipparcos subdwarf sample. These models give a closer match to the stars
on the unevolved main-sequence, with a typical offset of only $\sim 0.02$ magnitudes
in colour (or 0.1 to 0.15 in M$_V$). However, it is not possible to fit simultaneously 
both the colour at the turnoff (ages $\ge$ 16 Gyrs) and the luminosity of the
subgiant branch (ages $\le$ 14 Gyrs). 

We have also compared the Hipparcos-based subdwarf data to the sets of models
calculated recently by D'Antona, Caloi \& Mazzitelli (1997). These models
are ``classical'' in the sense that there are no specific abundance enhancements.
However, there are several significant differences in the theoretical framework
as compared with the BV92 or SC91 methodology. First, convection is treated using a
full scale turbulence analysis, rather than with the standard mixing-length approximation;
second, the effects of helium sedimentation are included; and, third, the models use
the equation of state developed by Rogers, Swenson \& Iglesias (1996). Taken together,
these three changes to the input physics in the models lead to both cooler temperatures
and lower luminosities for upper main-sequence stars of a given age.

DCM use the Kurucz (1993) (L, T$_{eff}$)  calibrations to transform 
the model isochrones to (M$_V$, (B-V)), and figure 12 plots the local
subdwarf data together with their results for abundances of [Fe/H]=-1.3, -1.5 and
-2.0 and ages of 10, 12 and 14 Gyrs. Of the three theoretical analyses discussed here,
these models provide the closest match to the main-sequence, turnoff and subgiant branch
data. The predicted giant branch is generally too red, but DCM point out
that, apart from dealing with the inherent physical uncertainties in modelling 
evolved stars, the predicted colours can be changed by -0.07 magnitudes simply
by adopting the Vandenberg (1992) temperature-colour relation rather than the
Kurucz calibration. In general, the isochrones predicted by these models are a
good match to the shape of the turnoff region in the colour-magnitude diagrams for 
M5, M13 and M92, and imply ages in the range 11 to 13 Gyrs.

None of the three models discussed here produces isochrones which are an
exact match to the shape of the cluster main sequences. Thus, there is
a subjective element involved in assessing which model is the closest match
to a given dataset. 
In comparing observational data to theoretical isochrones in the (M$_V$, (B-V))
plane, one tends to align the two datasets horizontally. This is misleading, 
since there is usually a factor of ten difference in the scales on 
the x- and y-axes, and what appears to be a
small offset in M$_V$ can be of considerably more significance than a
misalignment in (B-V). Moreover, the disparity in colour between the 
Kurucz and Vandenberg giant-branch calibrations underlines the continuing
problems in defining reliable colour-temperature relations. 
Bolometric corrections are generally better defined. 
Thus, matching the models to the data in luminosity/M$_V$, either using the
entire luminosity function or specific features, 
offers the prospect of a more robust age determination.

Neither SC91 nor DCM calculate theoretical luminosity functions, so we have
taken the absolute magnitude at the turnoff, M$_V$(TO), as our point of
reference. The cluster colour-magnitude diagrams are nearly vertical at
this point, so the apparent magnitude of the turnoff (the bluest point
on the main-sequence) is uncertain by at least $\pm 0.1$ magnitudes.
Figure 13 compares our estimates for each cluster against the predictions
of the three sets of models discussed here. Since neither the SC91 nor
DCM models allow for element enhancement, we have offset the data by
+0.15 dex in [Fe/H] when comparing the observations to these theoretical models.
In the case of the four clusters where the foreground reddening is
a source of significant uncertainty, our results are based on the value of
E$_{B-V}$ which aligns the subgiant branch with that of the appropriate
reference cluster (M13 or M92). Thus, we adopt the median values for the 
three metal-poor clusters and the higher reddening for NGC 6752. 
The open triangles indicate the position of the turnoff for the other
suggested values of the reddening towards those clusters.

There are two important conclusions to be drawn from the results presented
in figure 13. First, {\sl all} of the age estimates are 14 Gyrs or less -
well below the 15.8 Gyr average found by Bolte \& Hogan (1995). The
ages derived from the three sets of models for a given cluster (and reddening) 
also span a relatively small range, with typically only 1-2 Gyrs difference 
between the ages derived from the classical SC91 models and the (younger) BV92 
and DCM age estimates. Note, in particular, that if we adopt the HD 19445
calibration for the metal-poor clusters (i.e ($\delta (m-M)_0 = +0.12$ mag),
the average age increases by only $\sim 1$ Gyr: to 13 Gyrs if calibrated against
the SC91 models, and to 12 Gyrs in the case of the BV92 and DCM models.

Second, there is a suggestion of some dispersion in the cluster ages
amongst the metal-poor clusters. The age estimates for the
three clusters with low extinction, M5, M13 and M92, agree within the
1$\sigma$ uncertainties irrespective of the model one adopts. 
However,  if we adopt the standard values for the line-of-sight reddening towards M68 
and M15 (i.e. E${_B-V}$ = 0.07 and 0.11 magnitudes, 
corresponding to the uppermost open triangles plotted in
figure 13), then both clusters are significantly ($\sim 2$ Gyrs)
younger than M92. DCM arrive at similar conclusions based on their
isochrone-fitting of the M68 and M92 colour-magnitude diagrams.
 
The suggestion of an age-spread amongst globular clusters is 
by no means unprecedented: 
Bolte (1989) originally proposed that 
NGC 362 is 2-3 Gyrs younger than NGC 288, while Buonnano et al (1994)
have identified four  clusters in the outer halo
(Terzan 7, Arp 2, Pal 12 and Ruprecht 106) which appear to be 3-4 Gyrs. younger
than 47 Tucanae. All of these clusters have abundances [Fe/H] $> \sim -1.6$,
however, and, previous studies of the extreme metal-poor systems are
more equivocal. Chaboyer et al (1996) place M68 amongst the youngest Galactic globulars
based on the luminosity redward of the turnoff, on the subgiant branch.
These results are obviously dependent on the reddening that one adopts, 
and Vandenberg, Bolte \& Stetson (1990) have shown that the colour-difference
between the turnoff and the base of the giant branch, a reddening-independent
parameter, is identical within the observational uncertainties for M92 and M68.
On the other hand, the blue horizontal branch is more extended in both M15 
and M68 than in M92, suggesting a possible age difference, and Chaboyer, 
Demarque \& Sarajedini (1996) estimate an difference of 5 Gyrs between 
the ages of M92 and M68, with M15 having an intermediate age. 

In summary, depending on the reddening towards the individual clusters, 
our data {\sl may} indicate an age spread amongst the metal-poor halo
clusters. Irrespective of reddening uncertainties, however, it is clear
that the average age of even extreme metal-poor halo globular clusters 
is significantly less than 14 Gyrs, and may well be as low as 12 Gyrs. 

\subsubsection {Cluster ages and the early history of the Galaxy}

A younger Galactic halo has a number of significant
implications for models of Galactic evolution. First, and foremost, one is
not faced with the problem of explaining how the Galaxy managed to
form a substantial number of halo globular clusters, and perhaps most of the 
field stars in the halo, during a period of collapse very early in its history, 
and then allowed the star formation rate to dwindle to insignificant levels
for up to 4 Gyrs before embarking on the task of disk formation. The age of 
the Galactic disk is not yet well defined. However, Oswalt et al (1996) have set
a lower limit of 9.5 Gyrs, based on the shape of the white dwarf luminosity function
at low luminosities. A small number of open clusters, however, are suspected
of dating from the earliest epochs of disk formation, and 
Phelps (1997) has recently estimated an age of 12$^{+1}_{-2}$ Gyrs for 
Berkeley 17. This would make this -0.3 $\le$ [Fe/H] $\le$ 0.0 cluster a
close contemporary of both M5 and M92, systems with abundances lower by
almost a factor of 100. This suggests 
strongly that star formation started at approximately the
same time in {\sl both} the halo and disk stellar populations.

Our results also show weak evidence for a spread in ages amongst the 
metal-poor halo clusters, with the strongest support lying in Walker's (1994) thorough
analysis of the reddening towards M68. If confirmed by further observations, this 
would require the presence of metal-poor gas well after the first outburst of star 
formation throughout the Galaxy, and after several generations of supernovae and AGB 
stars have enriched the metal content of the interstellar medium. The most
likely mechanism for avoiding pollution by stellar ejecta is to form
any younger clusters from infalling primordial gas clouds. This is essentially a minor 
variation on the theme presented originally by Searle \& Zinn (1978).
Recent observational studies of distant halo stars, notably by Majewski, Munn \&
Hawley (1994), have provided evidence for kinematic sub-structure within the
halo, suggesting continued accretion of distinct systems. These results have
been discussed primarily in the context of mergers of relatively-massive,
dwarf spheroidal systems, and the consequent impact on the dynamical evolution 
of the disk. However, a significant fraction of the halo (which has a 
total mass only a few percent that of the disk) may well have been constructed over the initial
2-4 Gyr period of the Galaxy's lifetime by the accumulation of gas clouds with 
individual masses of no more than $\sim 10^6 M_\odot$. 
Continued accretion also provides a natural explanation for the
presence of two distinct kinematic components within the Galactic halo, 
as proposed by Hartwick (1987) and Sommer-Larsen \& Zhen (1990), and supported
by Majewski's (1992) detection of retrograde rotation in the outer halo.

As a qualitative outline of the overall formation scheme, we hypothesise that the
initial stages of Galactic formation occurred as a rapid collapse, in
the manner proposed originally by Eggen, Lynden-Bell and Sandage (1962).
During the first few $\times 10^8$ years, star formation occurred
throughout the proto-halo, with the lower angular-momentum processed
material falling deeper into the Galactic potential well and enriching
the proto-disk (and bulge). Most of the initial star clusters were
disrupted to form the field stars in the present Galactic halo.

The infalling gas collapses into a rotating disk, and star formation
commences within the disk no more than 0.5 Gyr after the first halo stars
formed. The likeliest candidates for first disk stars are the metal-weak 
"thick disk" stars identified by Morrison, Flynn \& Freeman (1990), and
it is likely that most of the globular clusters in Zinn's disk system
(Zinn, 1985) also formed during this phase of Galactic evolution. The
gas within the disk collapses further to form the thin disk - the precursor
to the component currently termed the old disk - with mild star formation
persists within "thick disk" material, whose larger scaleheight is maintained by 
continued infall of gaseous material. This accretion of 
primordial gas-clouds may also lead to the addition of 
younger, metal-poor globulars, such as M15 and M68, to the outer Galactic halo.

This is an hypothesis which clearly requires more quantitative development.
However, the characterisation of galaxy formation as a relatively unspectacular
process, involving no more than a moderate star formation rate extending over
several Gyrs, and with continued
accretion to relatively modest redshifts (z $\sim 2$), is in accord with
observations of faint galaxies in the Hubble deep field. In particular, the morphological
classification undertaken by van den Bergh et al (1996) shows a higher fraction
of merging/interacting systems and fewer grand design spirals than amongst
nearby galaxies.

\subsection {The extragalactic distance scale and the Hubble constant}

Given that we have proposed both a larger distance scale and a younger age
for the benchmark globular cluster system, it is incumbent upon us to
consider how these results impact on the investigations of the distance scale.
Since all current investigations of the Hubble expansion are tied, in some
respect, to the distance to the LMC, the net result of our proposed change in
the distance modulus is a decrease in the inferred Hubble constant.
The longer distance scale, favoured by Tammann \& Sandage (1996) is based on
a number of indicators, some of which are dependent on the assumed LMC distance, either
directly through Cepheid-based distances, as is the case for Type I supernovae
distance determinations (Sandage et al, 1996), or indirectly, through calibrations 
of the distance to M31 and M101. However, Sandage (1997, preprint) has argued
that the overall calibration is little-dependent on the distance assumed to the
LMC, and the inferred value of H$_0$ is unaltered at 55$\pm$5 km s$^{-1}$ Mpc$^{-1}$.
The implied cosmological time-scale remains consistent with the revised 
globular cluster ages presented here.

The expansion time-scale inferred from the short distance-scale
(H$_0 > 70$ km s$^{-1}$ Mpc$^{-1}$), however, is in
direct conflict with the $>15$ Gyr ages that had been assigned
previously to the globular cluster system. 
The distance to the LMC is the corner-stone of the HST Key Project,
since the observed Cepheid period-luminosity relations in external galaxies
are all scaled to match the LMC relation. Thus, increasing the distance modulus 
of the LMC from 18.5 (the value adopted in the analyses to date) to 18.65 increases 
the distance scale by $\sim 7 \%$. Moreover, since the results derived by
other methods, such as surface-brightness fluctuations (Tonry et al, 1997) and
the Tully Fisher method (Bureau, Mould \& Staveley-Smith, 1996), are also usually
tied to the Cepheid scale, the distance scale adopted in those surveys must be
increased by a similar factor. As an example of the resultant effect on the
derived H$_0$, the ``mid-term"  value of the Hubble constant presented by Freedman 
et al (1997), H$_0 = 73 \pm 10$ km s$^{-1}$ Mpc$^{-1}$, is reduced
to H$_0 = 68 \pm 9$ km s$^{-1}$ Mpc$^{-1}$. This corresponds to an expansion
age of 12-13 Gyrs for $\Omega_0$=0.2 and to 10 Gyrs for $\Omega_0$=1
(Freedman, priv. comm.). The former value is not inconsistent with 
the cluster age estimates discussed in the previous sections.

\section {Conclusions}

We have used high-accuracy parallax measurements made by the Hipparcos astrometric
satellite of nearby halo subdwarfs to refine the definition of the
main-sequence for metal-poor stars.
Combining these data with previous observations of
an additional three stars, we have utilised main-sequence fitting techniques to
re-determine the distances of three moderately metal-poor and four extremely
metal-poor globular clusters. We find that the distances to the former
clusters are underestimated by $\sim 2-5 \%$, and that the distances to
the extreme metal-poor clusters should be increased by $\sim 15 \%$. The
main source of the former scale-change rests with the revised zeropoint
provided by the Hipparcos parallax data; the larger change associated with
the metal-poor clusters reflects the fact that many previous distance
determinations were semi-empirical in nature, tied to matching theoretical
isochrones whose absolute zeropoint was set by one star, HD 103095. Our
new data show that the Bergbusch \& Vandenberg (1992) isochrones, used by
Bolte \& Hogan (1995) in their analysis of M92, are too blue in (B-V)
when matched against the subdwarf data. As a result, previously-derived
distances were too low and, as a consequence, age estimates too old.

Our revision of the distance scale to globular clusters has two main
consequences: first, the luminosity associated with horizontal branch stars 
is increased, particularly at the lowest abundances, leading to a
re-calibration of the RR Lyrae absolute magnitudes, and an inferred
distance modulus of 18.65 to the LMC. This brings the Cepheid and RR Lyrae 
scales into agreement. Second, the increased luminosity of
the main-sequence turnoff implies that the clusters are significantly
younger. Matching against evolutionary models calculated by Straniero 
\& Chieffi (1991), Bergbusch \& Vandenberg (1992) and D'Antona et al (1997), 
we derive average cluster ages of $\sim 11$ to 13 Gyrs. If the standard
reddening estimates for M68, M15 and M30 are correct, then these three clusters
are $\sim 2$ Gyrs younger than M92. However, decreasing the reddening by only 0.02
magnitudes brings the colour-magnitude diagrams for all four clusters into 
close alignment. 

Irrespective of the reddening uncertainties, our results imply a significantly
younger age for the clusters in the Galactic halo. The more conservative
Straniero \& Chieffi models indicate an age of $\sim 13$ Gyrs, while the 
other two models favour an age of only 12 Gyrs. Coupled with recent
results which push back the onset of star formation in the disk, this
suggests that the initial epochs of star formation in the disk and halo
were separated by only a few hundred million years. Finally, taken together
the increased distance scale and the
younger ages of the oldest globular clusters succeed in almost reconciling
the stellar and cosmological age estimates for the Universe.

\acknowledgments 
We acknowledge the substantial efforts made by the many people who have
made the Hipparcos project a striking success.
INR would also like to thank Wendy Freedman, Ian Thompson and David Hogg 
for commenting on earlier versions of this paper; Brian Chaboyer and
Francesca D'Antona for comments on theoretical models; Mike Bolte for
providing the list of subdwarf calibrators used previously to calibrate
the distance to M92;  Bob Hanson for further clarifying the
problems surrounding applying Lutz-Kelker corrections; and George Preston for 
throwing the reddening spitball.

\clearpage
\centerline{FIGURE CAPTIONS}
\vskip2em

\figcaption{Lutz Kelker corrections. The solid points mark the systematic
offset in M$_V$ as a function of $\sigma_\pi \over \pi$ calculated originally
by Lutz and Kelker and the solid line shows Smith's (1987) analytic
representation of these datapoints. The dotted, long-dashed and short-dashed
lines outline the corrections predicted by Hanson's formula for 
$n=2$, 3 and 4 respectively, where $n$ is the exponent of a power-law parallax
distribution. The $n=4$ (uniform density) case is equivalent to the original
Lutz-Kelker analysis.}

\figcaption{The lower panel shows a comparison between ground-based parallax 
measurements and Hipparcos astrometry for 26 stars in common between the current
sample and the Yale Parallax catalogue. The upper panel compares the
absolute magnitudes adopted by S96 for twelve of the stars used in their 
M5 distance determination against the Hipparcos data (solid dots) and the final YPC
data (triangles). The error bars in this diagram represent the 
combined formal uncertainties of the two M$_V$ estimates, and the
dotted line shows the mean offset between the S96 and Hipparcos calibration,
excluding HD 219617.}

\figcaption{ The main-sequence for metal-poor stars, defined by Lowell proper
motion stars with measured abundances and Hipparcos parallaxes with
precision of at least 12 \% }

\figcaption{Main-sequence fitting for the three intermediate-abundance
globular clusters discussed in this paper. 
The calibrating subdwarfs from the appropriate abundance range are plotted at 
the appropriate Lutz-Kelker
corrected absolute magnitudes, and the errorbars indicate 1$\sigma$ uncertainties.
The NGC 6752 fiducial sequence is plotted for the two reddening values listed in
Table 3.}

%fig 5

\figcaption{Main-sequence fitting for the four metal-poor globular clusters.
Again, the relevant set of calibrating subdwarfs is shown. The reddening toward
M15, M30 and M68 is not unambiguously determined, so we plot the results for
two values of the extinction for those clusters. The upper panel shows the 
best-fit calibration for the highest reddenings listed in Table 3, while the
lower panel plots the results if we adopt the intermediate values. The latter
lead to an almost exact alignment of the subgiant branch with the (low-reddening)
M92 data.}

%figure 6

\figcaption {The results of matching the M92 fiducial sequence to an
[Fe/H]=-2.1 mono-metallicity subdwarf sequence, using the Hipparcos stars as
reference and with the differential colour corrections taken from the BV92 models}

%figure 7

\figcaption {Main-sequence fitting for M92 using a subset of the S96 subdwarfs
as the local calibrators. The open squares mark the observed positions of those
stars {\sl at the Lutz-Kelker corrected absolute magnitudes derived by S96}.
The solid points mark the same stars after adjusting the (B-V) colours to
match an [Fe/H]=-2.1 isophote. The calibrated M92 fiducial sequence is plotted 
together with the BV92 [Fe/H]=-2.26 14 Gyr isochrone.}

% figure 8

\figcaption{A comparison in the (M$_V$, (B-V)) plane between the BV92 oxygen-enhanced 
isochrones and the location of the ten subdwarf calibrators available to
Bolte \& Hogan (1995). The solid points mark the actual colours and magnitudes
while the open circles show positions after adjusting to [Fe/H]=-2.26.
Each subdwarf is labelled with its metallicity, and the
three isochrones plotted for each abundance are for ages of 12, 14 and 16 Gyrs.
The M92 fiducial sequence is plotted for a distance modulus of 14.65 mag, as
derived by BH95}

% figure 9

\figcaption{A comparison between the BV92 theoretical isochrones and the
empirical main-sequence (M$_V$, (B-V)) defined by the local subdwarfs. Each
sub-panel plots 12, 14 and 16 Gyr isochrones for a given abundance, together
with the subdwarfs within the specified abundance range and the appropriate
globular cluster fiducial sequence. Note that the cluster data are matched to
the subdwarfs, not the isochrones.}

\figcaption {A comparison between the absolute magnitudes we derive for
intermediate and extreme metal-poor RR Lyraes based on the distance
moduli we derive respectively to M5 and the three clusters M92, M68 and M15.
Previously derived  (M$_V$, [Fe/H]) relations are also shown. The solid line marks the
results derived by Layden et al from statistical parallax analysis of the
local stars, while the dotted lines outline the 1$\sigma$ uncertainties. We
allow for overlap in abundance between their halo and "thick disk" samples.
The other linear relation plotted are from Carney et al (1992), Fernley (1994), 
Sandage (1993) and Feast (1997). The solid points mark our current results.}

% figure 11

\figcaption{A comparison between the SC91 theoretical isochrones and the
empirical main-sequence. As
in figure 9, we show a set of model isochrones, in this case for
ages of 10, 12, 14 and 16 Gyrs and abundances of [Fe/H]=-1.3, -1.7
and -2.0, together with the relevant subdwarf data and
a globular cluster fiducial sequence. The subdwarfs are drawn from the same
abundance ranges as in the corresponding panels in figure 9. Since these
models do not include any $\alpha$-element enhancement, we have not 
plotted the comparison with the [Fe/H]=-2.3 isochrones.}

\figcaption {A comparison between the DCM theoretical isochrones and the
empirical metal-poor main-sequence(s). The format is as in figures 9 and 11,
with the theoretical isochrones plotted for ages of 10, 12 and 14 Gyrs
and for abundances of [Fe/H] = -1.3, -1.5 and -2.0.}

% figure 13

\figcaption {Age estimation using the luminosity of the main-sequence turnoff.
The observed values of M$_V$(TO) are compared to the theoretical predictions of
the SC91, BV92 and DCM models. The dotted lines show the predicted variation of
M$_V$(TO) with abundance for a given age (listed on the right edge of each panel).
The solid points plot the observed values, with the open triangles showing the effect
of adopting the different estimates (given in Table 3) 
of the reddening towards M15, M68, M30 and NGC 6752. Since neither
the SC91 nor DCM models include element enhancements, we have offset the cluster data
by +0.15 dex in [Fe/H].}
 
\setcounter{figure}{0}
\begin{figure}
\plotone{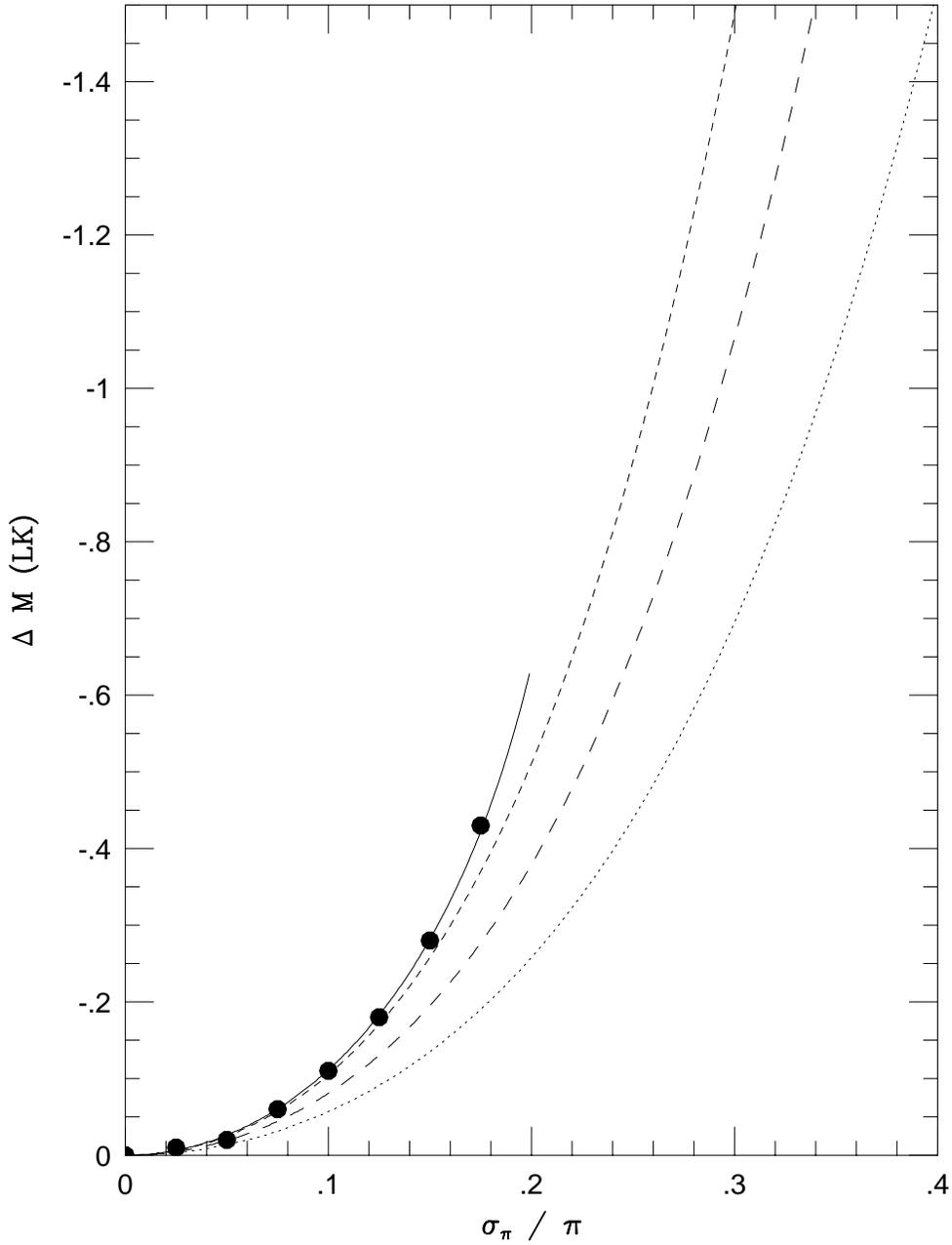}
\caption{Lutz Kelker corrections. The solid points mark the systematic
offset in M$_V$ as a function of $\sigma_\pi \over \pi$ calculated originally
by Lutz and Kelker. The dotted, long-dashed and short-dashed
lines outline the corrections predicted by Hanson's formula for 
n=2, 3 and 4 respectively, where n is the exponent of a power-law parallax
distribution.}
\end{figure}

\begin{figure}
\plotone{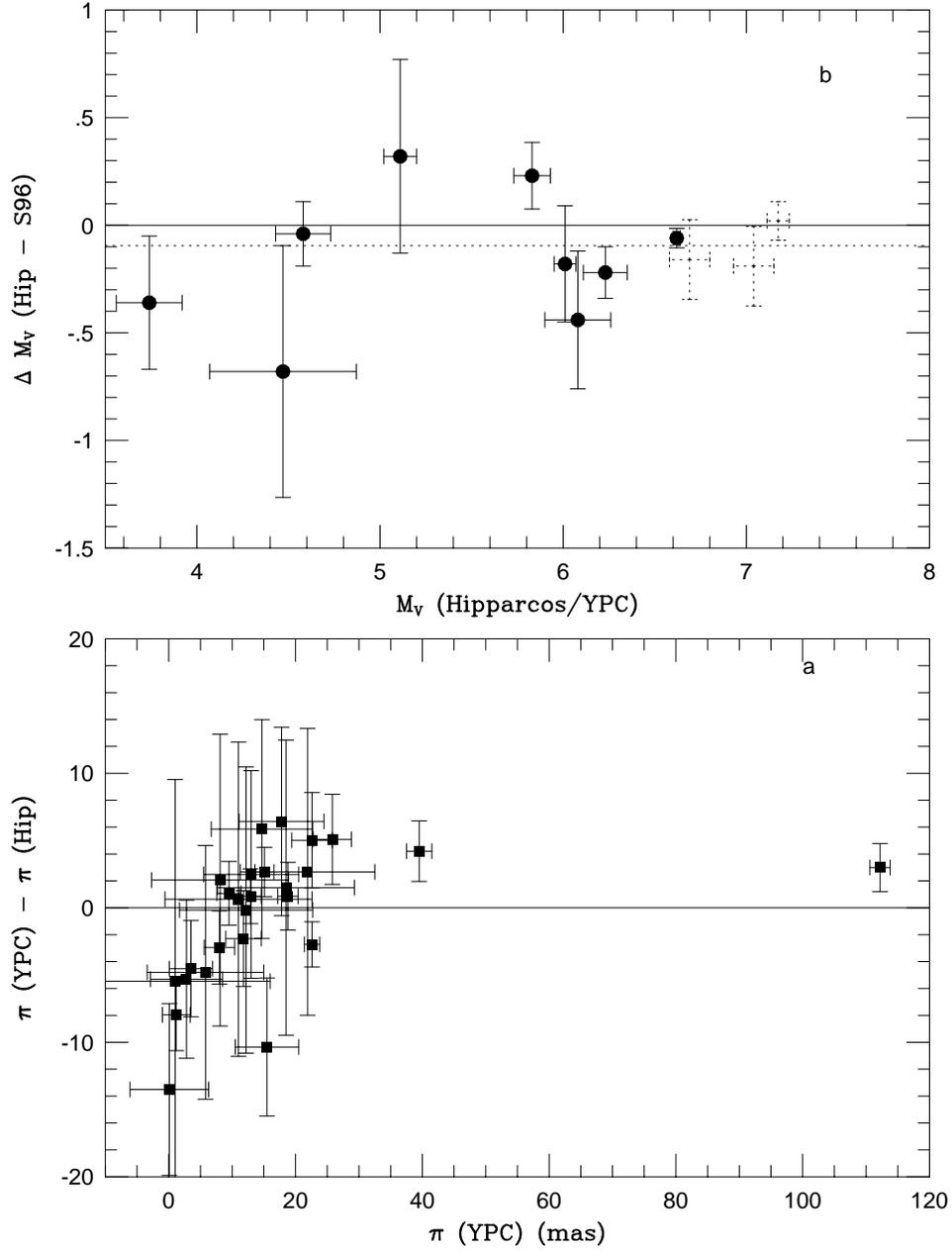}
\caption{The lower panel shows a comparison between ground-based parallax 
measurements and Hipparcos astrometry for 26 stars in common between the current
sample and the Yale Parallax catalogue. The upper panel compares the
absolute magnitudes adopted by S96 for twelve of the stars used in their 
M5 distance determination against the Hipparcos data (solid dots) and the final YPC
data (triangles). }
\end{figure}

\begin{figure}
\plotone{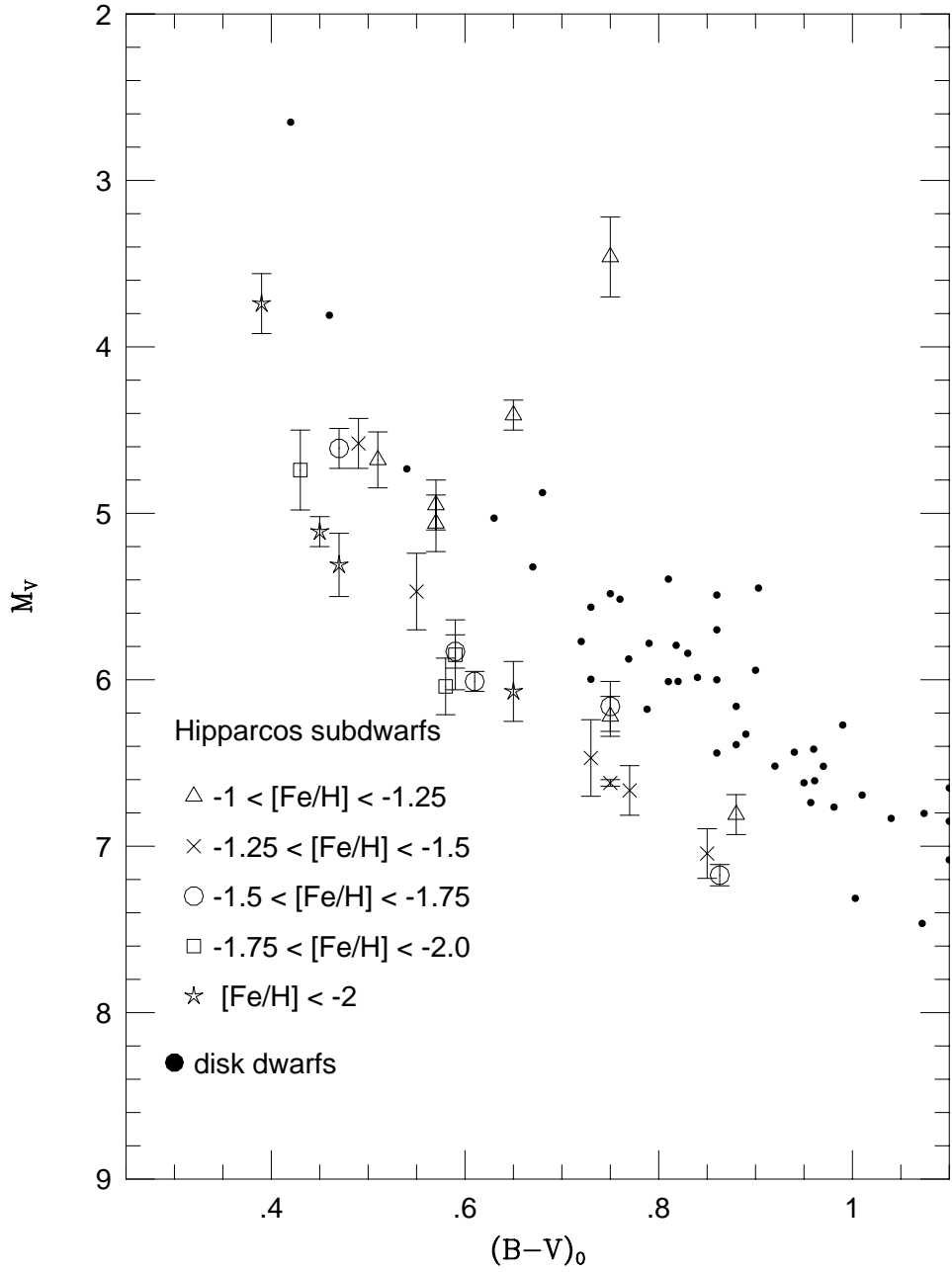}

\caption{ The main-sequence for metal-poor stars, defined by Lowell proper
motion stars with measured abundances and Hipparcos parallaxes with
precision of at least 12 \% }

\end{figure}

\begin{figure}
\plotone{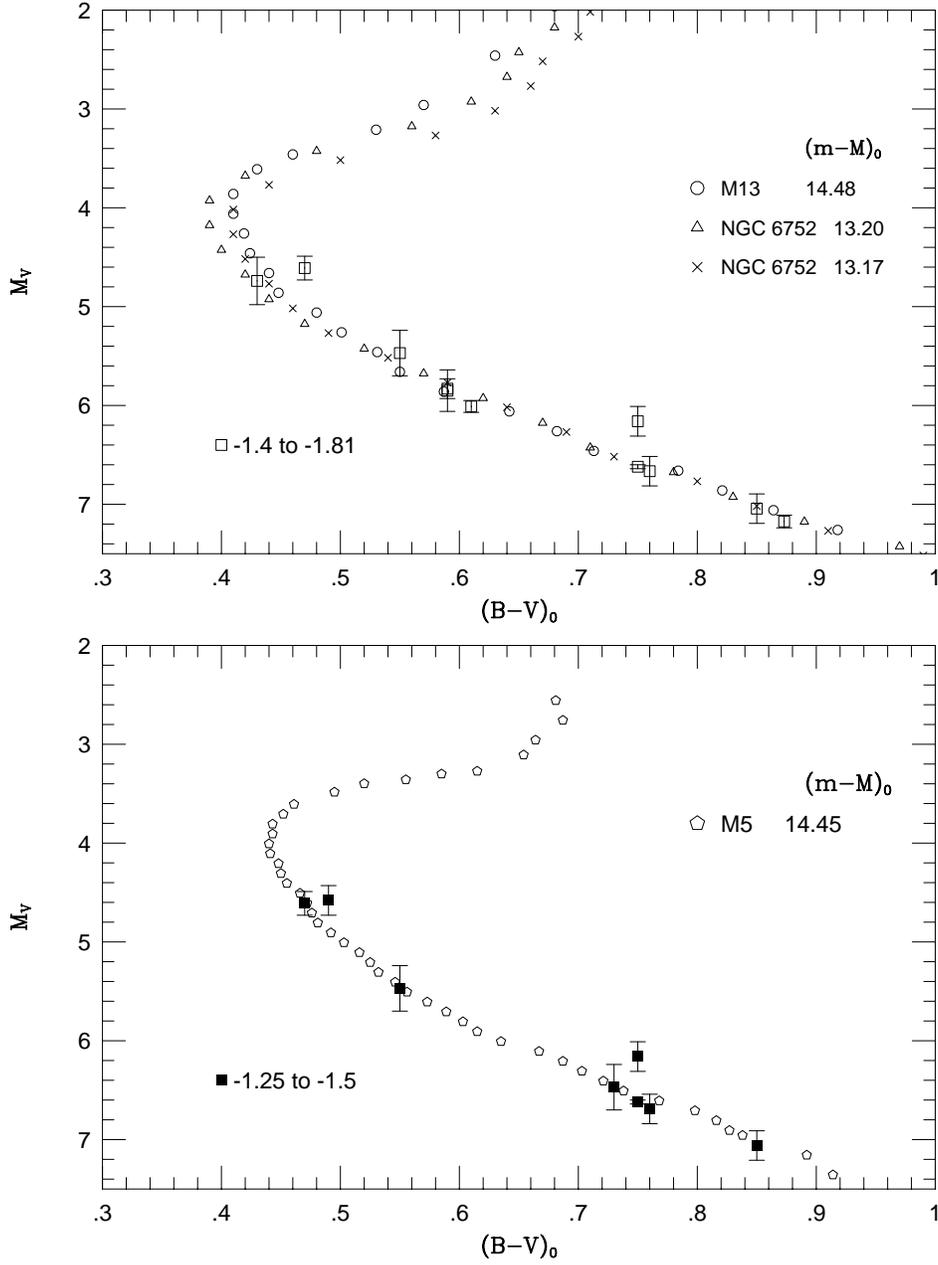}
\caption{Main-sequence fitting for the three intermediate-abundance
globular clusters discussed in this paper. 
The calibrating subdwarfs from the appropriate abundance range are plotted at 
the appropriate Lutz-Kelker
corrected absolute magnitudes, and the errorbars indicate 1$\sigma$ uncertainties.}
\end{figure}

\begin{figure}
\plotone{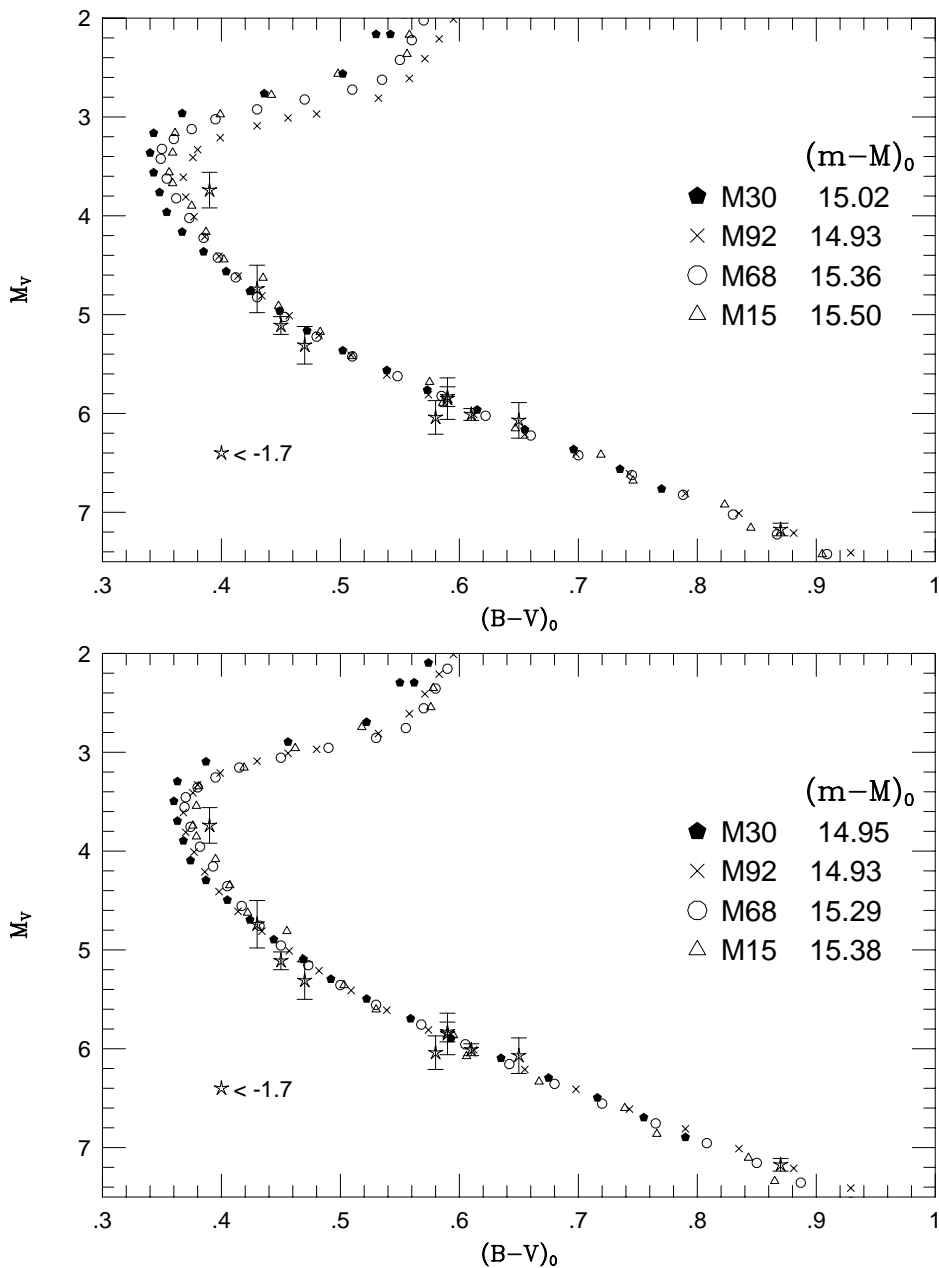}

%fig 5

\caption{Main-sequence fitting for the four metal-poor globular clusters.
Again, the relevant set of calibrating subdwarfs is shown. The reddening toward
M15, M30 and M68 is not unambiguously determined, so we plot the results for
two values of the extinction for those clusters. The upper panel shows the 
best-fit calibration for the highest reddenings listed in Table 3.}
\end{figure}

\begin{figure}
\plotone{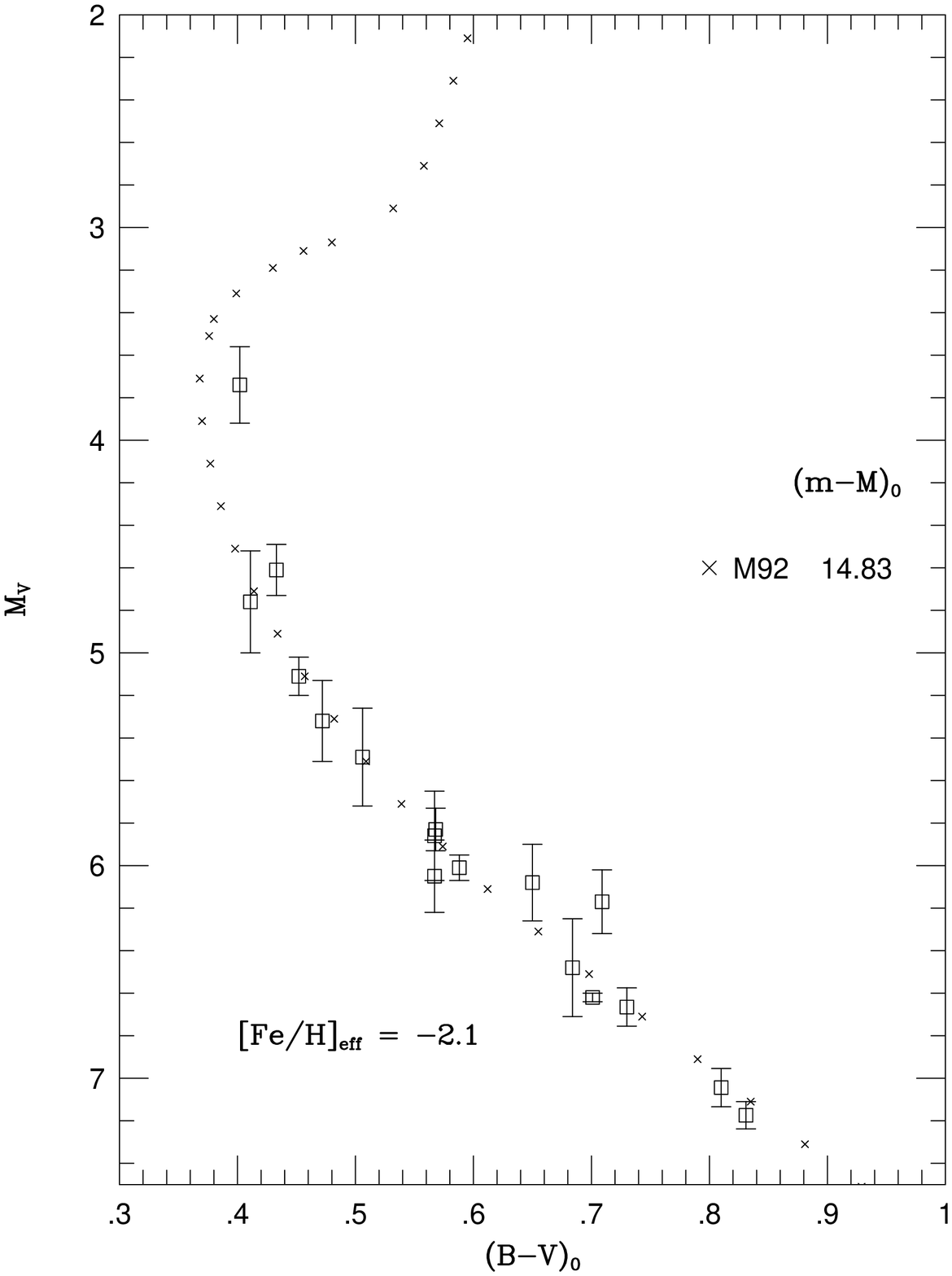}

%figure 6

\caption {The results of matching the M92 fiducial sequence to an
[Fe/H]=-2.1 mono-metallicity subdwarf sequence, using the Hipparcos stars as
reference and with the differential colour corrections taken from the BV92 models}

%figure 7
\end{figure}

\begin{figure}
\plotone{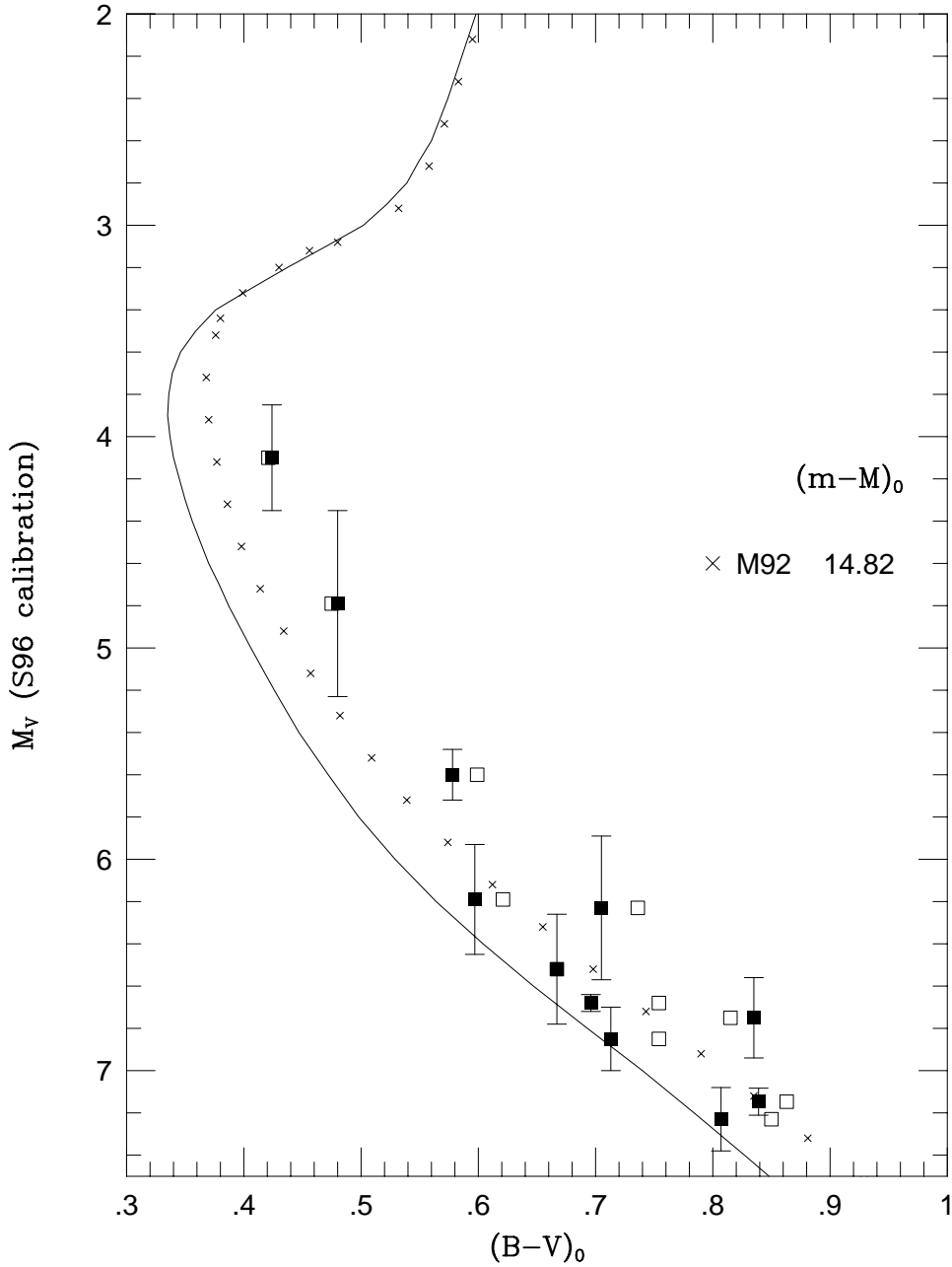}
\caption {Main-sequence fitting for M92 using a subset of the S96 subdwarfs
as the local calibrators. The open squares mark the observed positions of those
stars {\sl at the Lutz-Kelker corrected absolute magnitudes derived by S96}.
The solid points mark the same stars after adjusting the (B-V) colours to
match an [Fe/H]=-2.1 isophote. The calibrated M92 fiducial sequence is plotted 
together with the BV92 [Fe/H]=-2.26 14 Gyr isochrone.}

\end{figure}

\begin{figure}
\plotone{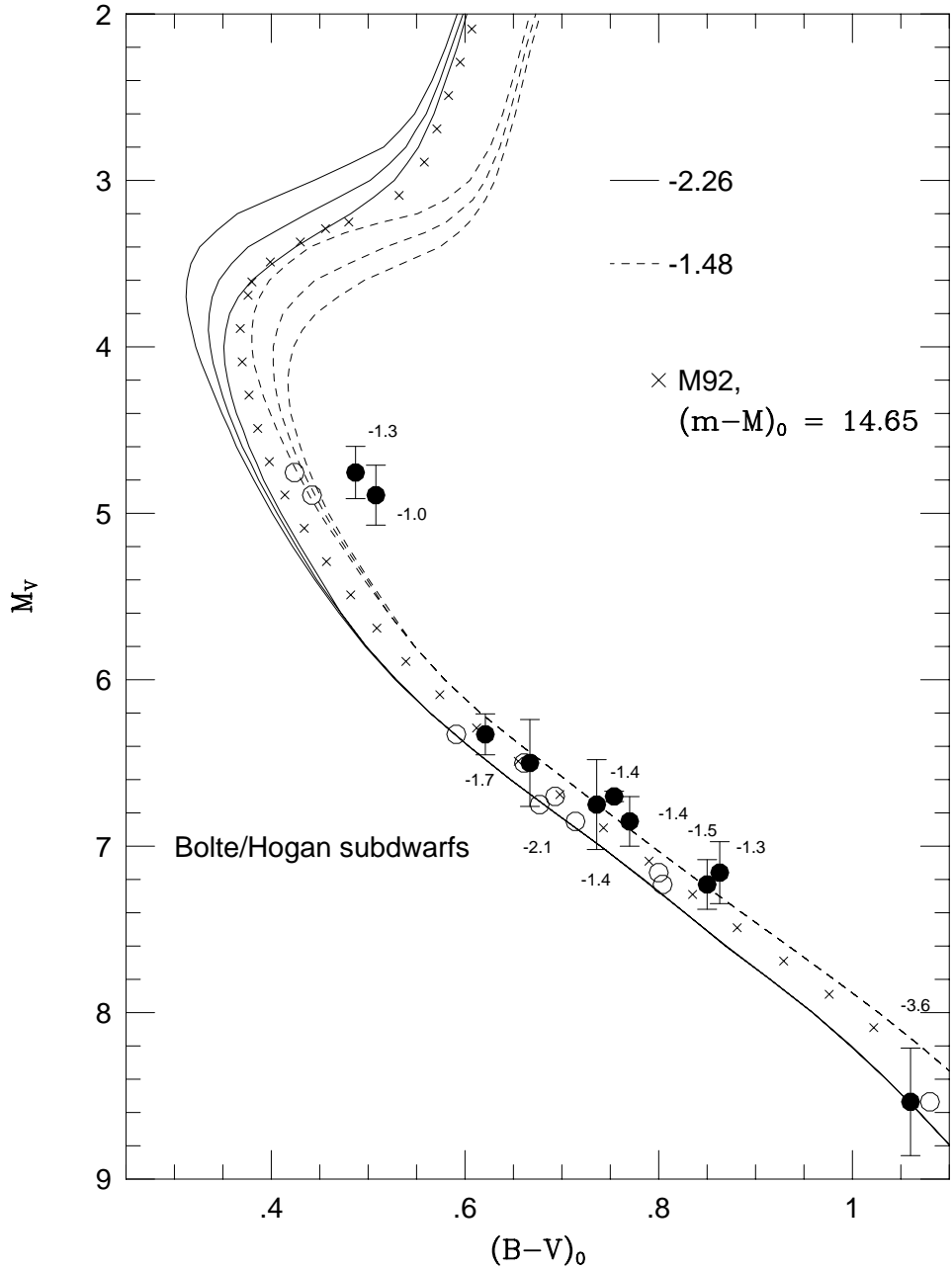}
% figure 8

\caption{A comparison in the (M$_V$, (B-V)) plane between the BV92 oxygen-enhanced 
isochrones and the location of the ten subdwarf calibrators available to
Bolte \& Hogan (1995). The solid points mark the actual colours and magnitudes
while the open circles show positions after adjusting to [Fe/H]=-2.26.
The three isochrones plotted for each abundance are for ages of 12, 14 and 16 Gyrs.}

\end{figure}

\begin{figure}
\plotone{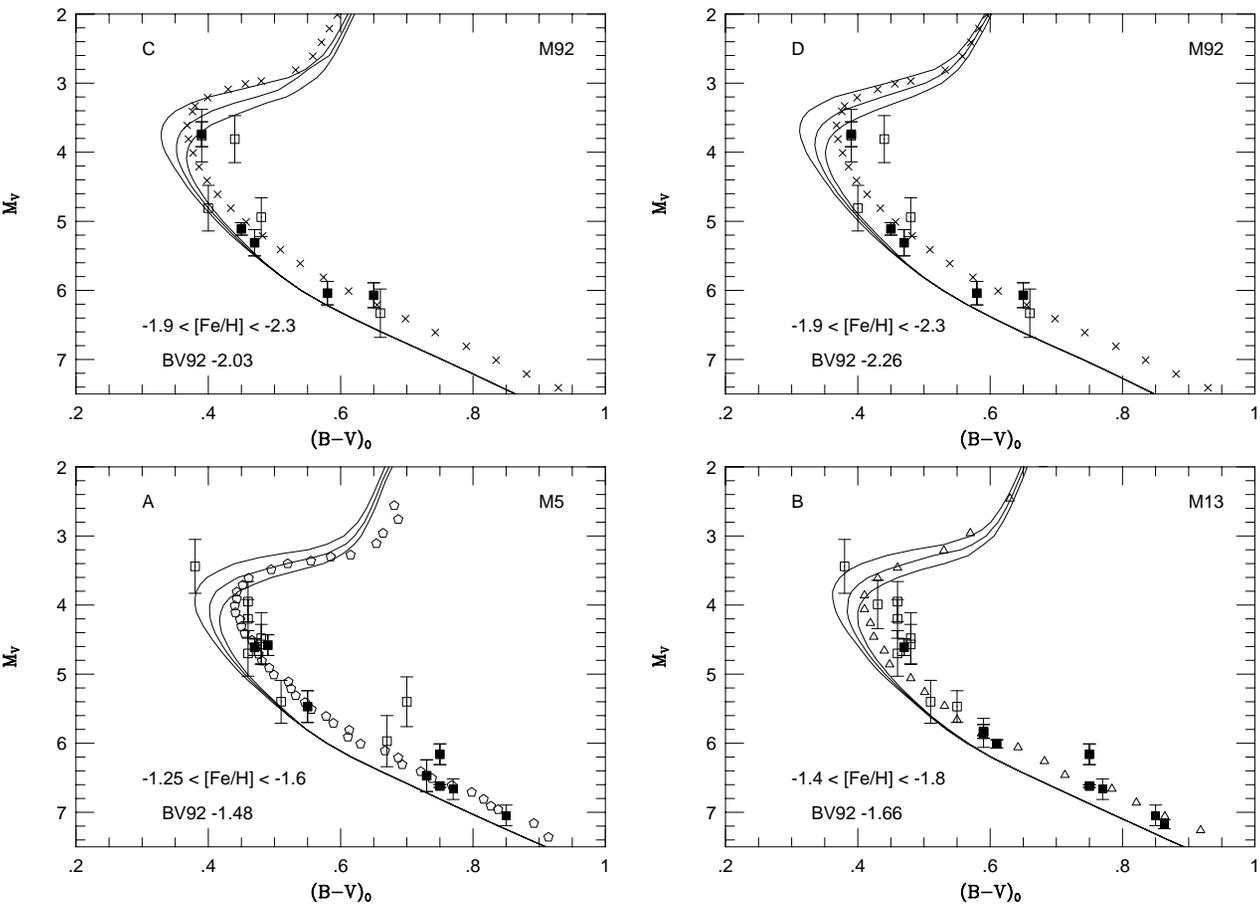}
% figure 9

\caption{A comparison between the BV92 theoretical isochrones and the
empirical main-sequence (M$_V$, (B-V)) defined by the local subdwarfs. Each
sub-panel plots 12, 14 and 16 Gyr isochrones for a given abundance, together
with the subdwarfs within the specified abundance range and the appropriate
globular cluster fiducial sequence. Note that the cluster data are matched to
the subdwarfs, not the isochrones.}

\end{figure}

\begin{figure}
\plotone{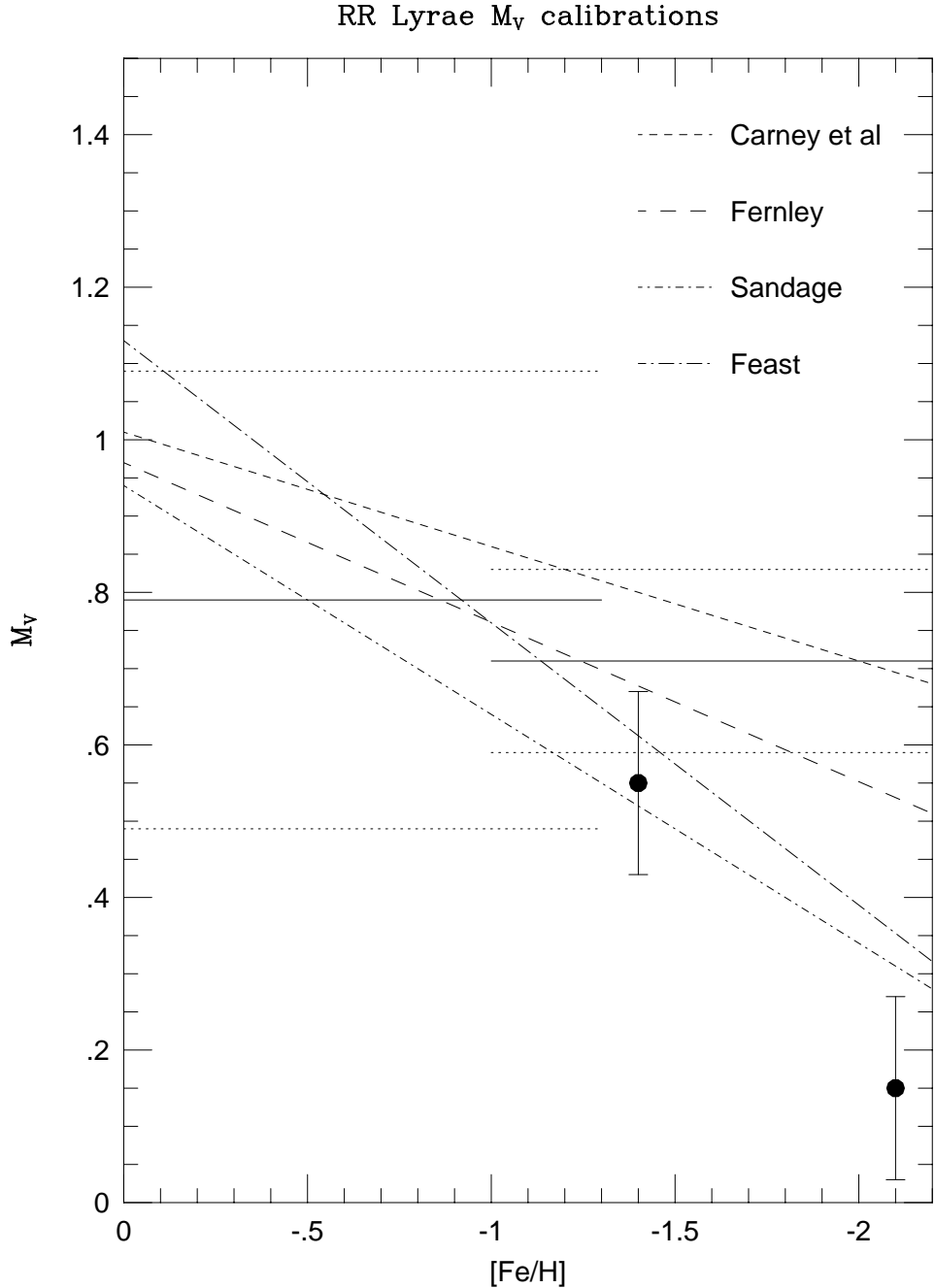}

\caption {A comparison between the absolute magnitudes we derive for
intermediate and extreme metal-poor RR Lyraes based on the distance
moduli we derive respectively to M5 and the three clusters M92, M68 and M15.
Previously derived  (M$_V$, [Fe/H]) are also shown. The solid line marks the
results derived by Layden et al from statistical parallax analysis of the
local stars. The solid points mark our current results.}

\end{figure}

\begin{figure}
\plotone{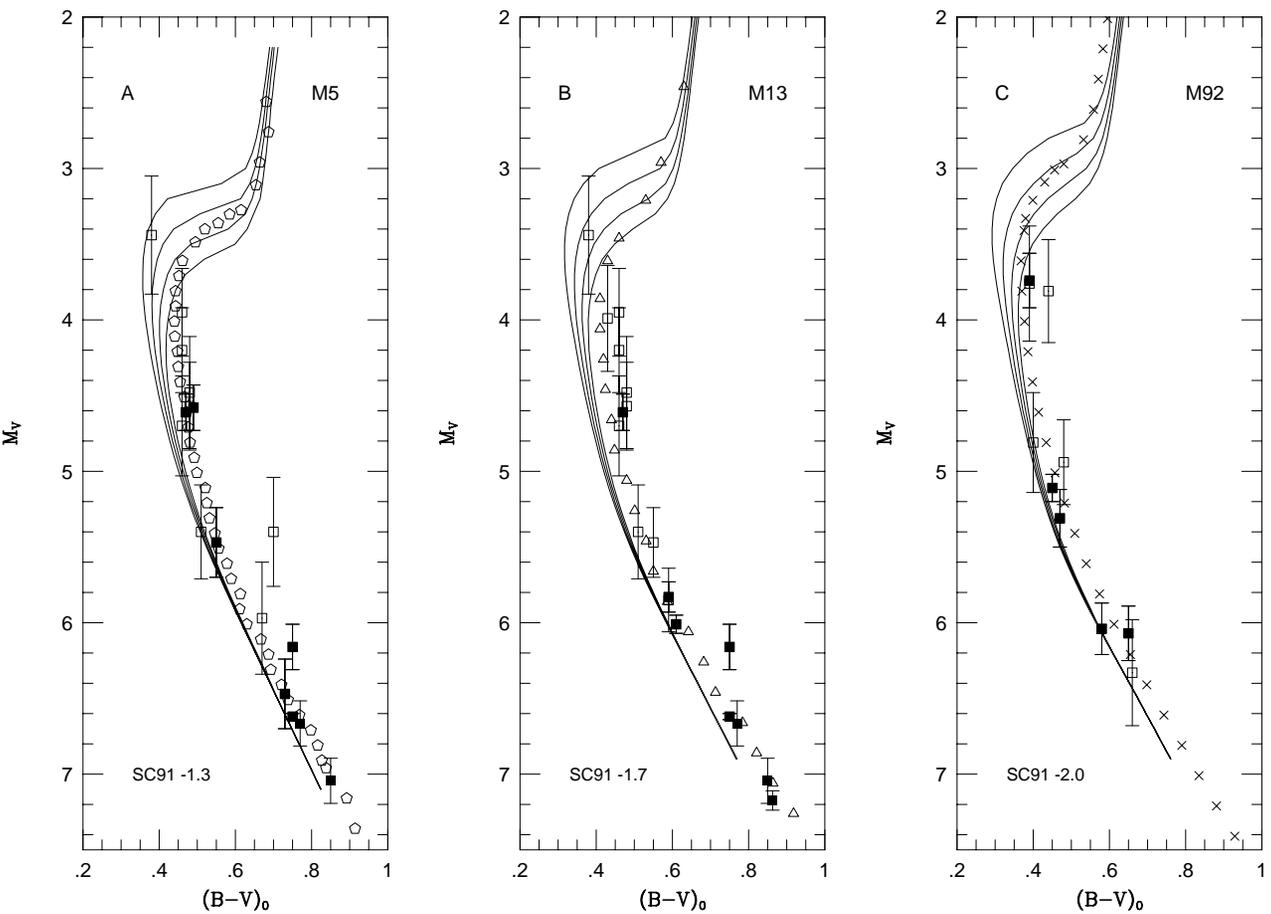}
% figure 11

\caption{A comparison between the SC91 theoretical isochrones and the
empirical main-sequence. As in figure 9, we show a set of model isochrones, for
ages of 10, 12, 14 and 16 Gyrs and abundances of [Fe/H]=-1.3, -1.7
and -2.0. }

\end{figure}

\begin{figure}
\plotone{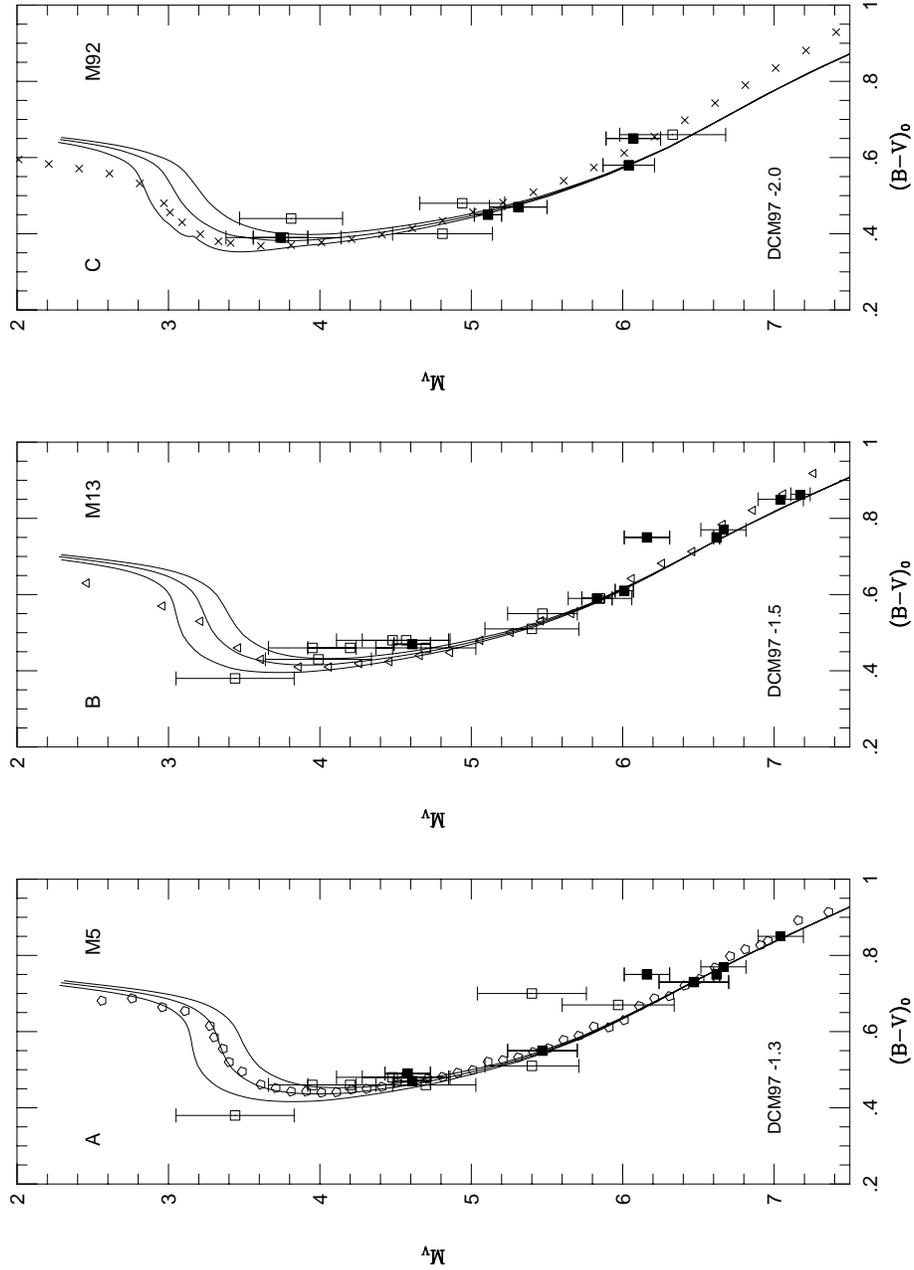}
\caption {A comparison between the DCM theoretical isochrones and the
empirical metal-poor main-sequence(s). The format is as in figures 9 and 11,
with the theoretical isochrones plotted for ages of 10, 12 and 14 Gyrs
and for abundances of [Fe/H] = -1.3, -1.5 and -2.0.}

\end{figure}

\begin{figure}
\plotone{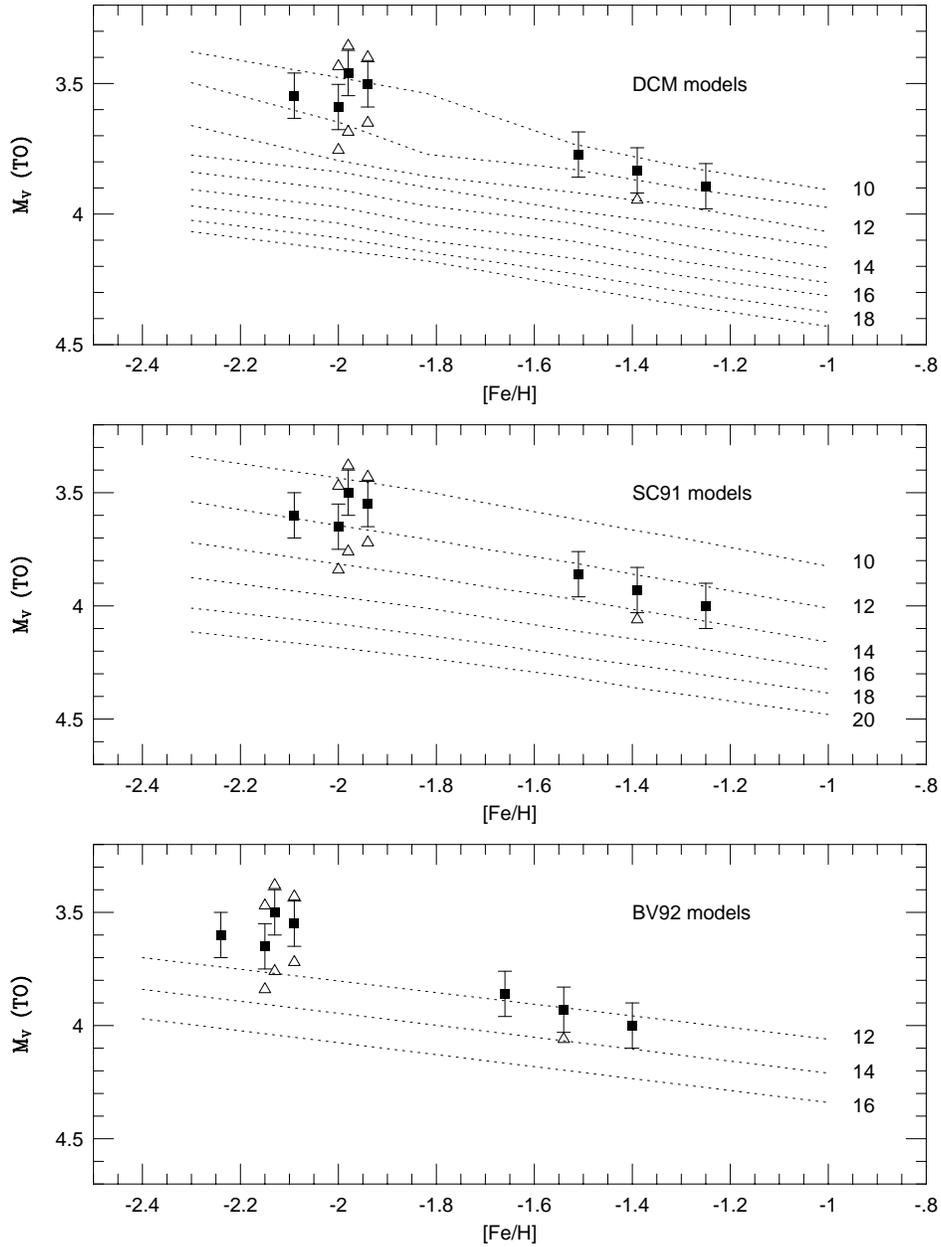}
% figure 13
\caption {Age estimation using the luminosity of the main-sequence turnoff.
The observed values of M$_V$(TO) are compared to the theoretical predictions of
the SC91, BV92 and DCM models. The dotted lines show the predicted variation of
M$_V$(TO) with abundance for a given age (listed on the right edge of each panel).}
\end{figure}
 
\end{document}